\documentclass[conference,compsoc]{IEEEtran}

\ifCLASSOPTIONcompsoc
  \usepackage[nocompress]{cite}
\else
  \usepackage{cite}
\fi
\usepackage{amsmath,amssymb}
\usepackage{graphicx}
\usepackage{booktabs}
\usepackage{array}
\usepackage{xcolor}
\usepackage{tikz}
\usepackage{acronym}
\usepackage[hyphens]{url}
\usepackage[hidelinks]{hyperref}
\usepackage[all]{hypcap}
\hypersetup{
  pdftitle={Fuel as Identity: Authenticating Inter-Satellite Laser Links after Total Key Compromise},
  pdfauthor={Leon Birkel, Mohammad Hamad, Sebastian Steinhorst}}

\graphicspath{{figures/}}

\makeatletter
\let\sp@saved@acro\acro
\let\acro\newacro
\acro{AEAD}[AEAD]{Authenticated Encryption with Associated Data}
\acro{AES}[AES]{Advanced Encryption Standard}
\acro{AoA}[AoA]{Angle of Arrival}
\acro{ASAT}[ASAT]{Anti-Satellite\acroextra{ (weapon)}}
\acro{BPSK}[BPSK]{Binary Phase-Shift Keying}
\acro{CCSDS}[CCSDS]{Consultative Committee for Space Data Systems}
\acro{CIA}[CIA]{Confidentiality, Integrity, and Availability}
\acro{COTS}[COTS]{commercial off-the-shelf}
\acro{DLR}[DLR]{German Aerospace Center}
\acro{ECC}[ECC]{Elliptic Curve Cryptography}
\acro{ECDLP}[ECDLP]{Elliptic Curve Discrete Logarithm Problem}
\acro{ESA}[ESA]{European Space Agency}
\acro{ESTOL}[ESTOL]{ESA Specification for Terabit/sec Optical Links}
\acro{FoV}[FoV]{Field of View}
\acro{FSO}[FSO]{free-space optical\acroextra{ (communication)}}
\acro{FWHM}[FWHM]{Full Width at Half Maximum}
\acro{GEO}[GEO]{Geostationary Earth Orbit}
\acro{GNSS}[GNSS]{Global Navigation Satellite System}
\acro{GPS}[GPS]{Global Positioning System}
\acro{HAPS}[HAPS]{High-Altitude Platform Station}
\acro{ISL}[ISL]{inter-satellite link}
\acro{IT}[IT]{Information-Theoretic}
\acro{KEM}[KEM]{key encapsulation mechanism}
\acro{LEO}[LEO]{Low Earth Orbit}
\acro{MEO}[MEO]{Medium Earth Orbit}
\acro{MitM}[MitM]{machine-in-the-middle}
\acro{OCT}[OCT]{Optical Communications Terminal}
\acro{OISL}[OISL]{optical inter-satellite link}
\acro{OOK}[OOK]{on-off keying}
\acro{OTP}[OTP]{One-Time Pad}
\acro{PAA}[PAA]{Point-Ahead Angle}
\acro{PAT}[PAT]{Pointing, Acquisition, and Tracking}
\acro{PHY}[PHY]{Physical Layer}
\acro{PKI}[PKI]{Public Key Infrastructure}
\acro{PLA}[PLA]{Physical-Layer Authentication}
\acro{PLS}[PLS]{Physical Layer Security}
\acro{PNT}[PNT]{Positioning, Navigation, and Timing}
\acro{QKD}[QKD]{Quantum Key Distribution}
\acro{RPO}[RPO]{Rendezvous and Proximity Operation}
\acro{RSA}[RSA]{Rivest--Shamir--Adleman\acroextra{ (cryptosystem)}}
\acro{RTT}[RTT]{Round-Trip Time}
\acro{SDA}[SDA]{Space Development Agency}
\acro{SGP4}[SGP4]{Simplified General Perturbations Model 4}
\acro{SNR}[SNR]{Signal-to-Noise Ratio}
\acro{SPARTA}[SPARTA]{Space Attack Research and Tactic Analysis}
\acro{SSA}[SSA]{Space Situational Awareness}
\acro{TDOA}[TDOA]{Time Difference of Arrival}
\acro{TLE}[TLE]{Two-Line Element\acroextra{ (set)}}
\acro{ToF}[ToF]{Time of Flight}
\acro{TUM}[TUM]{Technical University of Munich}

\let\acro\sp@saved@acro
\makeatother

\newtheorem{proposition}{Proposition}
\newtheorem{lemma}{Lemma}
\newtheorem{theorem}{Theorem}
\newtheorem{definition}{Definition}
\makeatletter
\let\sp@begintheorem\@begintheorem
\let\sp@opargbegintheorem\@opargbegintheorem
\def\@begintheorem#1#2{\sp@begintheorem{\phantomsection #1}{#2}}
\def\@opargbegintheorem#1#2#3{\sp@opargbegintheorem{\phantomsection #1}{#2}{#3}}
\makeatother

\newcommand{\dv}{\Delta v}
\newcommand{\Si}[1]{S_{#1}}
\newcommand{\KDF}{\mathrm{KDF}}
\newcommand{\gnat}{g_{\mathrm{nat}}}
\newcommand{\Tmax}{T_{\max}}
\newcommand{\linktarget}[1]{\raisebox{\normalbaselineskip}[0pt][0pt]{\hypertarget{#1}{}}}
\newcommand{\labellink}[3]{\hyperlink{#1:#3}{#2#3}}
\newcommand{\labeltarget}[3]{\linktarget{#1:#3}#2#3}
\DeclareRobustCommand{\asm}[1]{\labellink{asm}{A}{#1}}
\newcommand{\asmdef}[1]{\labeltarget{asm}{A}{#1}}
\DeclareRobustCommand{\limref}[1]{\labellink{lim}{L}{#1}}
\newcommand{\limdef}[1]{\labeltarget{lim}{L}{#1}}

\newcommand{\goaldef}[1]{\labeltarget{goal}{G}{#1}}
\DeclareRobustCommand{\rqref}[1]{\hyperref[sec:rq#1]{RQ#1}}
\DeclareRobustCommand{\symref}[2]{\hyperlink{sym:#1}{#2}}
\newcommand{\symdef}[2]{\linktarget{sym:#1}#2}
\newcommand{\pname}{\textsc{OrbitLock}}

\definecolor{PaletteBlue}{HTML}{0065BD}
\definecolor{PaletteDarkBlue}{HTML}{005293}
\definecolor{PaletteLightBlue}{HTML}{64A0C8}
\definecolor{PaletteLighterBlue}{HTML}{98C6EA}
\definecolor{PaletteOrange}{HTML}{E37222}
\definecolor{PaletteGreen}{HTML}{A2AD00}
\definecolor{PaletteGray}{HTML}{999999}
\definecolor{PaletteIvory}{HTML}{DAD7CB}
\definecolor{ink}{HTML}{1F1F1F}

\colorlet{cread}{PaletteBlue}     %
\colorlet{cinject}{PaletteOrange} %
\colorlet{cpaid}{PaletteGreen}    %
\colorlet{cbox}{PaletteIvory}     %
\colorlet{cmuted}{PaletteGray}    %

\colorlet{treadtext}{PaletteBlue}              %
\colorlet{tinjecttext}{PaletteOrange!75!black} %
\colorlet{tpaidtext}{PaletteGreen!62!black}    %

\newcommand{\tread}[1]{\textcolor{treadtext}{#1}}
\newcommand{\tinject}[1]{\textcolor{tinjecttext}{#1}}
\newcommand{\tpaid}[1]{\textcolor{tpaidtext}{#1}}
\newcommand{\badgebox}[1]{{\setlength{\fboxsep}{0.9pt}\colorbox{cbox}{\vphantom{Ag}\textbf{#1}}}}
\newcommand{\asmbadge}[1]{\badgebox{\asm{#1}}}
\newcommand{\asmbadgedef}[1]{\badgebox{\asmdef{#1}}}
\colorlet{calloutbar}{cmuted}
\colorlet{calloutfill}{cbox!45}
\newlength{\calloutpad}
\newlength{\calloutbarwidth}
\newlength{\calloutgap}
\newsavebox{\calloutbox}
\newcommand{\callout}[2]{\par\smallskip\noindent
  \sbox{\calloutbox}{\parbox{\dimexpr\columnwidth-2\calloutpad\relax}{\textbf{#1}\enspace#2}}%
  {\setlength{\fboxsep}{0pt}\colorbox{calloutfill}{%
    {\color{calloutbar}\vrule width\calloutbarwidth
      height\dimexpr\ht\calloutbox+\calloutpad\relax depth\dimexpr\dp\calloutbox+\calloutpad\relax}%
    \kern\calloutgap\usebox{\calloutbox}\kern\calloutgap}}%
  \par\smallskip}
\newcommand{\para}[1]{\paragraph{\textbf{#1}}}
\newsavebox{\yesbox}\newsavebox{\partlybox}\newsavebox{\nobox}
\sbox{\yesbox}{\tikz[baseline=-0.6ex]\fill[ink] (0,0) circle (2.3pt);}
\sbox{\partlybox}{\tikz[baseline=-0.6ex]{\draw[ink, line width=0.5pt] (0,0) circle (2.3pt); \fill[ink] (0,-2.3pt) arc (-90:90:2.3pt) -- cycle;}}
\sbox{\nobox}{\tikz[baseline=-0.6ex]\draw[ink, line width=0.5pt] (0,0) circle (2.3pt);}
\DeclareRobustCommand{\yes}{\usebox{\yesbox}}
\DeclareRobustCommand{\partly}{\usebox{\partlybox}}
\DeclareRobustCommand{\no}{\usebox{\nobox}}

\begin{document}

\title{Fuel as Identity:\\Authenticating Inter-Satellite Laser Links after Total Key Compromise}

\author{%
\IEEEauthorblockN{Leon Birkel}
\IEEEauthorblockA{Technical University of Munich\\
Munich, Germany\\
leon.birkel@tum.de}
\and
\IEEEauthorblockN{Mohammad Hamad\IEEEauthorrefmark{1}}
\IEEEauthorblockA{Technical University of Munich\\
Munich, Germany\\
mohammad.hamad@tum.de}
\and
\IEEEauthorblockN{Sebastian Steinhorst\IEEEauthorrefmark{1}}
\IEEEauthorblockA{Technical University of Munich\\
Munich, Germany\\
sebastian.steinhorst@tum.de}}

\maketitle

\makeatletter
\begin{NoHyper}
\long\def\@makefntext#1{\parindent 1em\indent#1}%
\footnotetext{\IEEEauthorrefmark{1}These authors jointly supervised this work.}%
\end{NoHyper}
\makeatother

\begin{abstract}
Optical inter-satellite links form the backbone of entire constellations in low Earth orbit, yet their security rests on keys alone. When an attacker obtains every key of a satellite, no secret is left to authenticate it again. Existing fixes fail, since ground rekeying encrypts new keys under the leaked ones, and post-compromise protocols assume an authenticated start. As a remedy, the satellite's orbit can take over as the trust anchor. The attacker then needs an impostor satellite that blocks and retransmits the beam, a chord beneath the curved orbit. Hence, the blocking point lies off free fall and costs the impostor continuous thrust. We formalize this position-gated adversary and prove, for satellites on a shared circular orbit, a tight lower bound on the propellant cost in delta-v of a point-mass craft that keeps blocking, however it moves. We present OrbitLock, a history-chained protocol that combines this position gate with ML-KEM. Each link authenticates on its own, and we sketch how combining links forces an impostor onto every link. Once the chain outlasts an impostor's propellant, every key back to the first is authenticated retroactively. A satellite thus needs no crosslink key before launch, so none can leak in the supply chain. The same mechanism heals the link after total key compromise, even without compromise awareness. A passive attacker is shut out at the next round, whereas an active one splits the chains and is exposed once its propellant runs out. The chain can then heal after an operator reset. As a result, every Starlink satellite has a partner on whose link a CubeSat-class impostor runs out of propellant in about two days. This holds against one rigid craft per link running the best attack we know, if each terminal enforces a 1 km exclusion zone, while the protocol only uses under 0.02\% of the link.
\end{abstract}

\section{Introduction}\label{sec:intro}

\Acfp{OISL} now route traffic across entire constellations. The public catalog lists more than 11{,}000 Starlink satellites in orbit, and recent versions carry laser terminals that move 100 to 200~Gbps over up to 5400~km~\cite{ali2026leotopology}. The \ac{SDA} has standardized an optical terminal so that spacecraft from different vendors can link to one another~\cite{sda_oct}.

\begin{figure}[t]
    \centering
    \includegraphics[width=\columnwidth]{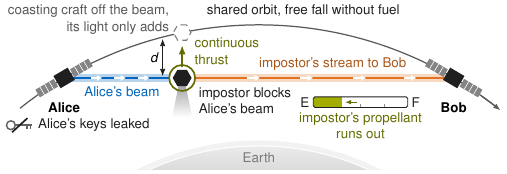}
    \caption{Fuel as identity: Alice and Bob share an orbit and a laser crosslink. Alice's keys have leaked, so Bob must authenticate her without them. An impostor must remove her signal at Bob. A craft that coasts on the orbit is off the beam, and its light only adds to hers. Therefore, the impostor must block her beam from a seat on the straight beam, a forced offset $\symref{d}{d}$ below the curved orbit, and thrust outward continuously. A CubeSat-class impostor 1~km from a terminal of a 1000~km link runs dry in a few days, while Alice and Bob coast for free. Not to scale.}
    \label{fig:hero}
\end{figure}

\para{Keys leak, and no key re-anchors them}
Quantum computers would break current public-key exchange and reveal every key it agreed on. Today, a flaw in a widely used commercial off-the-shelf communication module hands its telecommand signing key to an unauthenticated attacker~\cite{plantasatbleed}. Satellite firmware and telecommand interfaces have proven exploitable~\cite{willbold2023spaceodyssey}, and operational satellites still send sensitive data unencrypted~\cite{zhang2025don}. Closing such a flaw does not help, since every key the attacker read stays known to it. After a total key compromise, no one can visit the satellite to install a new secret. Nor can the operator certify a fresh key by radio, since the leaked keys let the attacker send its own. The satellite could confirm such a key by a maneuver that the operator tracks. Each such recovery costs propellant, and until the next one a leaked key is free to use.

\para{Relevance in practice}
Standard satellite key management cannot recover from a total key compromise. A satellite receives its first keys on the ground before launch, and every later key upload is encrypted under them~\cite{bader2024ttc}. An industry analysis of these \ac{CCSDS} procedures concludes that an attacker who learns all keys can never be shut out again, although no mission can exclude such a leak~\cite{bader2024ttc}.
Existing solutions do not close this gap. Post-compromise protocols such as the double ratchet heal only from an authenticated start, and only once the attacker turns passive~\cite{perrin2016doubleratchet, cohngordon2016pcs}. On a network, it never has to. Quantum key distribution needs an authenticated classical channel~\cite{scarani_security_2009}. Physical-layer authentication of satellites is statistical~\cite{topal_physical_2022, abdelsalam_ml-driven_2025, smailes2023satiq}.

\para{The adversary already flies}
Inspector spacecraft already shadow and spy on foreign satellites. Russia's Luch/Olymp-K parked near more than two dozen geostationary communication satellites for weeks to months, typically inside their uplink beams~\cite{swf2026counterspace}. France called its approach to Athena-Fidus an act of espionage~\cite{defensenews2018parly}, and its successor reportedly came within about 1~km of Intelsat 10-02 in January 2025~\cite{csis2025sta}. Russia and China test such approaches in low Earth orbit as well~\cite{swf2026counterspace}. Such a craft needs only a telescope and a laser to attack a crosslink. Laser-based constellations, therefore, face this adversary today. We target low Earth orbit, where these constellations fly and where the attacker's cost is about 1400 times higher than in geostationary orbit at equal link length.

\begin{figure*}[t]
    \centering
    \includegraphics[width=\textwidth]{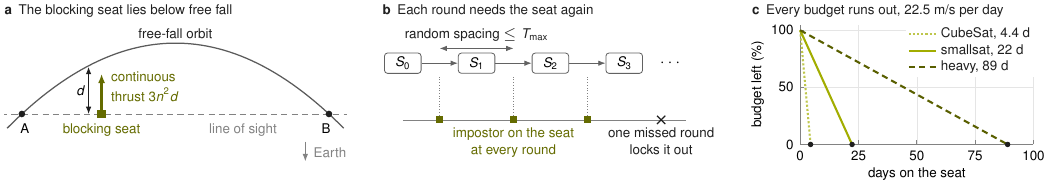}
    \caption{The argument of this paper in three panels, for a shared circular orbit at 550~km. (a)~The line of sight between Alice (A) and Bob (B) is a straight chord, and the orbit is an arc above it. An impostor must block Alice's beam, so its seat lies on the chord, a forced offset $\symref{d}{d}$ below free fall, and must keep thrusting outward to stay there. (b)~\pname{} advances a shared secret $\symref{Si}{S_i}$ in rounds at random times. Every round requires an impostor to occupy that seat again, and an impostor that misses one round can never form the next secret. (c)~The thrust drains every propellant budget. Beside a terminal on a 1000~km link, the seat costs 22.5~m/s of $\symref{dv}{\Delta v}$ per day, so a CubeSat-class budget of 100~m/s lasts 4.4 days and a 2~km/s platform 89 days. Geometry exaggerated.}
    \label{fig:teaser}
\end{figure*}

\para{Insight}
The orbit is a trust anchor that survives the leak of every key. The argument is a chain of four steps (Figures~\ref{fig:hero} and~\ref{fig:teaser}). First, an impostor that poses as Alice must remove her signal at her partner Bob, or her stream and its own add up at his detector. Second, light from off the line of sight cannot cancel her signal, because independent light adds in intensity at a direct-detection receiver. Third, the impostor must block her beam from a seat on the line of sight. Fourth, that line is a straight chord below the curved orbit, so the seat lies below free fall and needs continuous outward thrust. A passive reader of both directions also sits below free fall, but it gains nothing while ML-KEM holds. We state the thrust in $\symref{dv}{\dv}$, the velocity change that propulsion delivers, so a bound tells an operator which fixed budgets afford an attack and for how long.

\para{Closest prior work}
A recent poster abstract~\cite{birkel2026poster} stated this insight and mapped the Dolev--Yao~\cite{dolevyao1983} capabilities to seats. It priced a hovering attacker on three reference links, proposed a 1~km exclusion zone, and sketched a key chain. It gave no security definition, proofs, full protocol, or evaluation on a real constellation.

\para{Approach}
We build \pname{} on this physical gate. Alice and Bob run rounds at random times, at most 10~s apart. Each round mixes fresh random bits, which only a receiver inside both beams observes, with a fresh ML-KEM key~\cite{fips203}. As in the double ratchet of Signal~\cite{perrin2016doubleratchet}, every round authenticates under the secret of the previous one. The ML-KEM key never crosses the link, so a passive attack is never enough, even from inside both beams. Therefore, an impostor must substitute shares, which on a laser link needs a \tpaid{paid seat} on the line of sight at every round. Breaking the cryptography spares it only the rounds before it entered. Each link authenticates on its own, and we sketch how to combine links. Position verification is impossible against collusion, so we allow one rigid craft per link. The cost grows with the standoff each terminal enforces, here 1~km.

\para{Contributions}
This paper makes five contributions.
\begin{itemize}
    \item \textbf{A security definition that prices position.} We introduce budget-conditioned authentication, to our knowledge the first security experiment that bounds an attacker's advantage by the propellant it spends. Position-based cryptography asks where a prover is, and we ask what it costs to stay there. Because off-axis light cannot erase the genuine signal, modification needs the on-axis seat (Sections~\ref{sec:threat} and~\ref{sec:experiment}).
    \item \textbf{A tight floor on a shared orbit.} We prove that a point-mass craft holding a seat pays at least $3n^2 d\,T$, however it moves. Hovering attains the bound, and drifting, coasting, or leaving between rounds saves at most a fixed amount or a small fraction. Numerical arguments extend the floor beyond the shared orbit (Sections~\ref{sec:gate} and~\ref{sec:general}).
    \item \textbf{A quantified attack surface.} We size every cone from measured flight hardware, including a secrecy cone and the acquisition phase, and show that for a Gaussian beam no free seat reads both directions on a link that decides a recovery. The zone that keeps the attacker off the cheapest seats is thus only about 440~m$^3$ per terminal, in a known direction (Sections~\ref{sec:gate} and~\ref{sec:checks}).
    \item \textbf{\pname, specified in full.} We give the full protocol, with fail-closed acquisition, retries, an independent chain per link, and in-band checks that need no secret, among them a direction check that defeats tracking capture. Staying accepted costs an impostor $\symref{dv}{\dv}$ even if the cryptography fails, and while ML-KEM and the authenticated encryption hold, lasting acceptance means that it paid since the leak. The chain heals against a passive attacker after total key compromise, even without compromise awareness (Section~\ref{sec:protocol}).
    \item \textbf{Evaluation on a real constellation.} We price 17{,}433 possible Starlink crosslinks from public orbital data. With one attacking craft per link and a 1~km exclusion zone, about 99 percent of satellites have a link that exhausts a CubeSat-class craft within a week. The floor matches propagation with Earth's oblateness ($\symref{J2}{J_2}$) and drag to within 0.25 percent and NASA station-keeping practice to about 3 percent, and a released artifact recomputes every bound. Electric propulsion enlarges only the budget (Section~\ref{sec:eval}).
\end{itemize}

\section{Background}\label{sec:background}

\para{Laser terminals}
An \ac{OISL} terminal points a narrow beam at its partner and receives through a narrow field of view. Our reference is CubeISL, the crosslink version of the smallest laser terminal flown, the CubeLCT of the \ac{DLR}~\cite{osiris4cubesat_2024, cubelct_2023, dlr_cubeisl} (Table~\ref{tab:ref}). We choose it because its wide cones favor the attacker most. Commercial terminals with 80 to 100~mm apertures have beams about ten times narrower~\cite{tesat_scot80}, which barely moves the costs below but shrinks the zone a terminal must watch. DLR measured a half-angle of 102~$\mu$rad on the 20~mm aperture of the CubeLCT during integration, and this terminal then flew on the PIXL-1 CubeSat~\cite{roediger2025osiris4cubesat}. We credit the terminal with 1~W at 1550~nm, the output of DLR's amplified extension~\cite{roediger2020o4c}, since more power also favors the attacker. It uses \ac{OOK} with direct detection, which the \ac{SDA} standard requires terminals to support~\cite{sda_oct}.

\para{Three link phases}
During acquisition, the terminal scans an uncertainty cone set by the open-loop pointing error of its bus, about $\pm1^\circ$ for a common CubeSat. During tracking, a closed loop keeps the partner inside the 1.3~mrad field of view of its four-quadrant sensor~\cite{rueddenklau2024acquisition}. Acquisition matters only at round zero, because later rounds authenticate under the chain, and there a direction check and the fail-closed rule turn a second source into a visible outage (Sections~\ref{sec:checks} and~\ref{sec:freeseats}). Tracking costs the attacker about as much as communication (Table~\ref{tab:cones}), so we price the communication phase.

\begin{table}[t]
    \centering
    \caption{Reference parameters. The terminal is CubeISL, the attacker classes span flown hardware, and the orbit is the shared running example.}
    \label{tab:ref}
    \footnotesize
    \begin{tabular}{@{}l r@{}}
        \toprule
        \multicolumn{2}{@{}l}{\emph{Terminal and protocol}} \\
        Aperture, divergence half-angle $\symref{theta0}{\theta_0}$ & 20~mm, 102~$\mu$rad \\
        Transmit power, wavelength & 1~W, 1550~nm \\
        Modulation, data rate & \ac{OOK}, 100~Mbps \\
        Tracking field of view, range & 1.3~mrad, 1500~km \\
        Acquisition cone, open-loop pointing & $\pm$17.45~mrad \\
        Largest round spacing $\symref{Tmax}{\Tmax}$ & 10~s \\
        \midrule
        \multicolumn{2}{@{}l}{\emph{Orbit}} \\
        Altitude, radius $\symref{R}{R}$ & 550~km, 6921~km \\
        Mean motion $\symref{n}{n}$, period & 1.1~mrad/s, 96~min \\
        \midrule
        \multicolumn{2}{@{}l}{\emph{Attacker classes, budget $\symref{dvmax}{\dv_{\max}}$}} \\
        CubeSat, smallsat, heavy platform & 100~m/s, 500~m/s, 2~km/s \\
        \bottomrule
    \end{tabular}
\end{table}

\para{Orbital motion}
A lower orbit is a faster orbit. A craft that wants to stay below another craft without drifting ahead must keep pushing outward, against the stronger pull of gravity at its lower altitude. The push grows with the offset, and every cost in this paper follows from it~\cite[Ch.~7]{curtis2020orbital}.

\begin{figure}[t]
    \centering
    \includegraphics[width=\columnwidth]{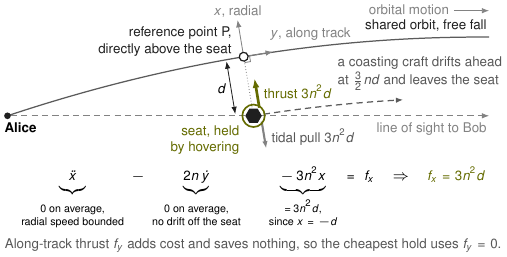}
    \caption{Why an attacker that must stay on a seat pays exactly $\symref{holdrate}{3n^2 d}$. In the frame that moves with the shared orbit, Alice, the seat, and the reference point P directly above the seat are at rest. At the forced offset $\symref{d}{d}$ below P, the tidal term of~\eqref{eq:cw} pulls with $\symref{holdrate}{3n^2 d}$. A coasting craft lets the Coriolis term carry this load, but it drifts ahead and leaves the seat. A craft that stays has no mean radial acceleration and no mean along-track motion, so only the tidal term remains, and radial thrust pays it most cheaply. Not to scale.}
    \label{fig:hover}
\end{figure}

Formally, a circular orbit of radius $\symref{R}{R}$ has mean motion $n = \sqrt{\mu/R^3}$. Motion near it follows the Clohessy--Wiltshire equations~\cite{curtis2020orbital}. With $\symref{x}{x}$ radial, $\symref{y}{y}$ along-track, and $f$ the thrust per unit mass,
\begin{equation}
    \ddot{x} - 2n\dot{y} - 3n^2 x = f_x, \qquad
    \ddot{y} + 2n\dot{x} = f_y.
    \label{eq:cw}
\end{equation}
The origin of this frame is the reference point directly above the craft, which moves with the orbit and is not Alice. A craft held at radial offset $\symref{d}{d}$ below the reference point with no along-track motion needs the continuous radial thrust $\symref{holdrate}{3n^2 d}$. A craft on a free, slightly lower orbit keeps its offset for free but drifts ahead at about $\tfrac{3}{2}nd$. Figure~\ref{fig:hover} shows why a staying attacker pays exactly this term. The spent $\symref{dv}{\dv}$ is the time integral of the thrust magnitude and fixes the propellant through the rocket equation.

\para{Attacker budgets}
A spacecraft leaves the ground with a fixed $\symref{dv}{\dv}$ budget. We price three classes, and each matches spacecraft that fly today. The CubeSat class of 100~m/s matches the three 13.4~kg HawkEye~360 Pathfinder microsatellites with 96~m/s~\cite{nasa2026soaprop}. The smallsat class of 500~m/s matches the Mira transfer vehicle, which offers 500~m/s at its full 300~kg payload and has maneuvered close to a second Mira~\cite{nasa2026soaild}, and True Anomaly's Jackal, rated up to 800~m/s~\cite{trueanomaly2025jackal}, which chased an evading target for the US Space Force in 2026~\cite{jewett2026victushaze}. The heavy class of 2~km/s matches a geostationary satellite of 1.5 to 2~t, which needs 1.8 to 2.35~km/s~\cite{oleson1997geo, glogowski2019servicing}. A hold cost in m/s per day thus converts into days of attack at a glance.

\section{Threat Model}\label{sec:threat}

\para{System and setting}
Two satellites of a constellation, Alice and Bob, communicate over one \ac{OISL} and carry the reference terminal. As the running example, they share one circular orbit at 550~km, and Section~\ref{sec:general} treats other near-circular geometries. Alice has suffered a total key compromise. The adversary has read every secret that Alice stores. It never ran code on Alice or Bob and cannot command them. No secret authenticates Alice anymore. Bob, her neighbor in the constellation, must authenticate Alice before the constellation trusts her with new keys. We analyze the raw signal of one link, since a property that holds without encryption still holds with it. Each of Alice's other links runs the same protocol and authenticates her on its own.

\para{Design goals}
The goal is an authentication primitive with three properties. (\goaldef{1})~It needs no secret of Alice that outlives the leak and no notice of it. (\goaldef{2})~The longer an impostor stays accepted, the more it pays, without an upper limit. Therefore, every fixed budget runs out, and we can compute when. (\goaldef{3})~It uses a negligible share of the link and, in its baseline form, needs no hardware beyond two star tracker heads, so satellites in orbit that meet the requirements of its checks can adopt it by a software update.

\para{Security goals and non-goals}
Authenticity is the primary goal, as it can be used to establish confidentiality. Availability, the identity of the operator, and implants that survive a reset are non-goals. Cryptography never provided availability, and far cheaper denials exist, up to destroying the satellite. The ground segment and supply-chain implants are out of scope.

\para{Adversary}
The adversary is a third party with at most one rigid maneuvering craft per link of Alice over the whole attack (\asm{9} in Table~\ref{tab:asm}). Each craft has a budget $\symref{dvmax}{\dv_{\max}}$, which counts any propellant it receives in orbit. It goes beyond the closest standard adversaries in three ways (Section~\ref{sec:related}). It holds every key of Alice, never has to turn passive, and may act and hold the line of sight with any aperture and receiver. The Dolev--Yao attacker, the adversary of post-compromise security, and the wiretap eavesdropper each lack one of these. Any Kepler orbit at any phase is free, and only the continuous $\symref{dv}{\dv}$ for a forced position is priced. Tethers and booms are out of scope (\asm{15}). In low Earth orbit, inspectors shadowed foreign satellites from tens of kilometers~\cite{swf2026counterspace}, far outside the zone of \asm{14}.

\para{Position-gated Dolev--Yao attacker}
We gate the Dolev--Yao attacker, who reads, modifies, injects, replays, and drops any message, by physics. Each capability on an \ac{OISL} has a positional precondition. \tread{Intercepting} breaks confidentiality and needs a receiver inside the beam. \tinject{Spoofing} and \tinject{injecting} break authenticity and integrity and need a transmitter inside the receiver's field of view. \tpaid{Modifying} breaks integrity and needs the on-axis seat, where the attacker blocks the genuine beam. The attacker holds a capability only inside the matching region and has no channel access outside all of them. Replay combines a read seat with a later inject seat, and drop survives only as the blocking half of a modification.

\para{No erasure}
Off the line of sight, the attacker can add light at a direct-detection receiver but cannot remove the genuine signal, because independent fields add in intensity~\cite{saleh2019photonics}. Canceling the signal would need a phase-locked, mode-matched copy held stable to tens of nanometers while the craft move and vibrate. Integrity codes rest on the same property~\cite{capkun2008integritycodes}, and the radio attacks that overcome it work at centimeter wavelengths~\cite{poepper2011manipulations}. Modification with chosen content is, therefore, confined to the \tpaid{on-axis seat}, once the checks of Section~\ref{sec:freeseats} expose a free craft that overpowers the genuine signal or steers a beam.

\para{Assumptions}
Table~\ref{tab:asm} lists all assumptions. Where a modeling choice is open, we take the option that favors the attacker (\asm{8}), so every bound on authentication is a floor.

\begin{table}[t]
    \centering
    \caption{Assumptions, grouped by what they constrain. The beige labels mark \asm{9}, \asm{14}, and \asm{15}, the assumptions on which the headline numbers rest. Section~\ref{sec:discussion} discusses relaxing each. The results also need a transmitting Alice (\asm{2}) and the tracking of the operator (\asm{6}). In the PDF, each label links to its row.}
    \label{tab:asm}
    \footnotesize
    \begin{tabular}{@{}l p{7.4cm}@{}}
        \toprule
        \multicolumn{2}{@{}l}{\emph{Trust and information}} \\
        \asmdef{1} & No key, certificate, or other secret trust anchor authenticates Alice. Public data, the signed ephemeris and time, is available. The only identity anchor is the trajectory. \\
        \asmdef{2} & Alice and Bob are genuine, and Alice keeps transmitting to Bob. The adversary can read Alice's secrets, not Bob's, and cannot run code on them or command them, for example because they check commands against an operator public key. \\
        \asmdef{3} & Following Kerckhoffs's principle~\cite{kerckhoffs1883}, the adversary knows everything except fresh secrets. This includes the protocol, the terminal design, and the public ephemeris of both endpoints. Both endpoints draw fresh secrets from a physical random source in every round, so no leaked state predicts them. \\
        \asmdef{6} & Both terminals point beaconlessly from ephemeris, time, and attitude, as in the \ac{SDA} \ac{OCT} standard. The operator signs each ephemeris, accurate to about 10~m, from tracking that does not trust the satellite's own reports. The public catalog is accurate only to about a kilometer at epoch~\cite{vallado2006spacetrack}. Time comes from an onboard clock that the operator checks to a millisecond, attitude from two star tracker heads, and neither from satellite navigation, which a nearby craft can spoof. \\
        \midrule
        \multicolumn{2}{@{}l}{\emph{Channel physics}} \\
        \asmdef{4} & The link runs in vacuum with negligible scattering. \\
        \asmdef{5} & The transmit beam is Gaussian with the measured divergence of Table~\ref{tab:ref}, which already includes the beam-quality widening ($M^2 = 1.2$), and without side lobes. This favors the defender. \\
        \asmdef{7} & The waveform is \ac{OOK} with direct detection. \\
        \asmdef{13} & No erasure. Off the line of sight, the adversary can add to but not lower the genuine signal at a linear direct-detection receiver. \\
        \midrule
        \multicolumn{2}{@{}l}{\emph{Adversary}} \\
        \asmdef{8} & Every open modeling choice but \asm{5} favors the attacker. \\
        \asmbadgedef{9} & At most one adversary craft ever attacks each link of Alice. \\
        \asmdef{11} & The adversary holds each capability only inside its region. \\
        \asmdef{12} & Any Kepler orbit and phase is free. Thrust pays every forced, non-Keplerian hold, against a budget $\symref{dvmax}{\dv_{\max}}$ that includes refueling. \\
        \asmbadgedef{15} & Each adversary craft is one rigid body, with no tether or boom. \\
        \midrule
        \multicolumn{2}{@{}l}{\emph{Operations}} \\
        \asmdef{10} & The adversary reaches and holds its position without being tracked from the ground. \\
        \asmbadgedef{14} & Each terminal keeps its exclusion zone, the first kilometer of its tracking field of view, clear of intruders with its own sensing (Section~\ref{sec:checks}). \\
        \bottomrule
    \end{tabular}
\end{table}

\para{What the primitive proves}
The primitive proves that the spacecraft at the far end follows the trajectory its public orbit predicts, round after round, and that no impostor can take its place on the link without paying. It does not prove which craft flies that trajectory or which operator controls it. It is, therefore, a kinematic continuity primitive. It bootstraps or complements conventional authentication after a key leak but does not replace it. A recovery by maneuver complements it.

\section{Overview}\label{sec:overview}

\pname{} layers a physical gate under standard cryptography.

\para{Two gates}
Each round draws on two independent gates. The physical gate prices, in $\symref{dv}{\dv}$, the position needed to observe or change both link directions. The cryptographic gate makes a passive attack useless, since the ML-KEM secret key never leaves Alice, so an impostor must turn active. An impostor that stays accepted defeats the physical gate at every round, even with broken cryptography, and while the cryptography holds, at every round since the leak.

\para{Two phases}
Round zero has no history yet. The fail-closed acquisition and the checks that need no secret guard it, but an attacker already on the on-axis seat passes it, so the $\symref{dv}{\dv}$ that accumulates later certifies round zero. From round one on, each round authenticates under the previous secret. A key leak adds no phase but restarts the paid history. An attacker with the leaked secret that misses one round is locked out, and one that leaves after substituting shares has split the two chains, so failures call the operator. The defender runs the same protocol throughout and need not know whether a key has leaked.

\section{The Physical Gate}\label{sec:gate}

The seat that blocks the beam lies below free fall, as does every seat that sees both link directions, and holding it costs at least $\symref{holdrate}{3n^2 d}$ of $\symref{dv}{\dv}$ per unit time. This section derives the cones, the forced offset $\symref{d}{d}$, and that floor.

\subsection{Where Each Capability Can Be Exercised}

\para{Read cone}
An eavesdropper can demodulate only inside the transmit beam. We grant it an unbounded telescope and the most sensitive receiver demonstrated, about 0.069 photons per bit~\cite{dacha2025record}. This figure comes from pulse-position modulation and over-credits an \ac{OOK} attacker (\asm{8}). The Gaussian beam of \asm{5} carries a fraction $\exp(-2\theta^2/\theta_0^2)$ of the transmitted power $P_{\mathrm{tx}}$ beyond angle $\symref{theta}{\theta}$ from its axis. The readable angle solves $P_{\mathrm{tx}}\exp(-2\theta^2/\theta_0^2) = N_b R_b\,hc/\lambda$ with $N_b = 0.069$ and $R_b = 100$~Mbps, which gives 0.380~mrad. A margin of 0.21~mrad for pointing errors, the mean tracking error of 71~$\mu$rad plus its $3\sigma$ deviation of 140~$\mu$rad that the terminal showed in orbit~\cite{rueddenklau2024acquisition}, yields the \tread{read cone} of half-angle 0.59~mrad. Outside it, no aperture collects enough light for a single bit. A stricter \emph{secrecy cone} caps the leak itself. The Holevo bound~\cite{holevo1973} limits what any measurement extracts from the two coherent states of \ac{OOK}, and a leak of $10^{-9}$ bits per transmitted bit gives a half-angle of 0.71~mrad. We price the read cone, since no built receiver reaches the Holevo bound, and the secrecy cone would lower the cost of the cheapest reading seat by under half a percent.

\para{Inject cone}
Injection and spoofing require a transmitter inside the receiver's field of view, a cone with its apex at the receiver. A seat in this \tinject{inject cone} is necessary but not sufficient. Alice keeps transmitting (\asm{2}), and with direct detection, added light cannot erase hers (\asm{13}). Like modification, injection with chosen content thus needs the on-axis seat (Section~\ref{sec:freeseats}). The cone has a half-angle of 0.65~mrad in tracking, set by the 1.3~mrad four-quadrant sensor~\cite{rueddenklau2024acquisition}, and up to 17.45~mrad in acquisition (Table~\ref{tab:cones}).

\begin{table}[t]
    \centering
    \caption{The cones that gate each capability. A cone contains the shared orbit only within the free stretch $2R\theta$ of the opposite terminal, so free seats serve one direction only. The last two columns give the forced offset $\symref{d}{d}$ and the hold cost of a seat inside the cone 1~km from the terminal at its tip, on the 1000~km link. A seat that serves both directions must pay them.}
    \label{tab:cones}
    \footnotesize
    \setlength{\tabcolsep}{3.5pt}
    \begin{tabular}{@{}l r r r r@{}}
        \toprule
        Cone & Half-angle & $2R\theta$ & $\symref{d}{d}$ & m/s per day \\
        \midrule
        \tread{Read}, demodulation & 0.59~mrad & 8.2~km & 71.6~m & 22.3 \\
        \tread{Read}, secrecy $10^{-9}$ & 0.71~mrad & 9.9~km & 71.5~m & 22.3 \\
        \tinject{Inject}, tracking & 0.65~mrad & 9.0~km & 71.5~m & 22.3 \\
        \tinject{Inject}, acquisition & 17.45~mrad & 242~km & 54.7~m & 17.1 \\
        \tpaid{Modify}, on axis & 0 & none & 72.2~m & 22.5 \\
        \bottomrule
    \end{tabular}
\end{table}

\para{The chord lies below the orbit}
All costs follow from one fact. The line of sight is a straight chord, the shared orbit is an arc above it, and the two meet only at the terminals. At distance $\symref{X}{X}$ from Alice on a link of length $\symref{L}{L}$, free fall passes a radial gap
\begin{equation}
    \gnat(X,L) = \frac{X(L-X)}{2R}
    \label{eq:gap}
\end{equation}
above the chord. At $X = 1$~km, the gap is 36.0, 72.2, and 108~m for $\symref{L}{L}$ of 500, 1000, and 1500~km.

\para{Forced offset}
Inside a cone of half-angle $\symref{theta}{\theta}$, the attacker can rise at most $\theta X$ toward the orbit. It rises that far, because every meter closer to the orbit lowers its cost. Therefore, it keeps the \tpaid{forced offset}
\begin{equation}
    d(X,L,\theta) = \gnat(X,L) - \theta X.
    \label{eq:d}
\end{equation}
The offset is positive at every priced seat, so every such seat costs $\symref{dv}{\dv}$. The cheapest point is the upper edge of the cone at the smallest distance $\symref{X}{X}$ the attacker may reach (\asm{14}). In the read cone, the attacker gains only 0.59~m against a 36~m gap at $X = 1$~km, so $\symref{d}{d}$ is essentially the full gap. Even the acquisition cone leaves $d = 18.6$~m on the 500~km link.

\begin{figure}[t]
    \centering
    \includegraphics[width=\columnwidth]{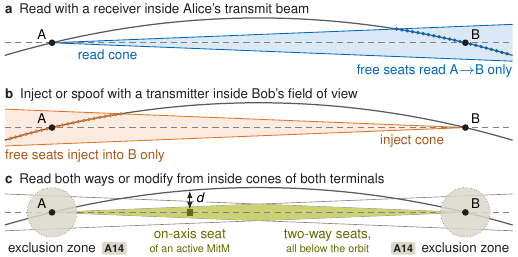}
    \caption{Where each capability can be exercised, in the frame that co-rotates with the shared orbit (angles and orbit bulge exaggerated). (a)~Reading needs a receiver inside Alice's transmit beam, the blue read cone. (b)~Injecting or spoofing needs a transmitter inside Bob's receiver field of view, the orange inject cone. Each cone contains the orbit only near the terminal it points at, so the free seats there, drawn dotted, serve one direction only. (c)~Reading both directions needs a seat inside cones from both terminals, and an active \acf{MitM} must sit on the line of sight itself, the on-axis seat. These green two-way seats lie below the orbit everywhere between the terminals. Each terminal keeps its exclusion zone clear (\asm{14}), which needs to cover only the first kilometer of its own field of view.}
    \label{fig:zones}
\end{figure}

\para{Free seats serve one direction}
A cone of half-angle $\symref{theta}{\theta}$ with its apex at one terminal contains the shared orbit only within $2R\theta$ of the other terminal (Figure~\ref{fig:zones}a,b). A craft co-orbiting there reads one direction or adds light to it at no cost. Being inside both cones at once requires both distances to stay below $2R\theta$. No point meets this on a link longer than $4R\theta$, which is 16.4~km for the read cone and 483~km for the acquisition cone. A link that decides a recovery is far longer than 16.4~km, so no free seat reads both directions there (\asm{5}). Shorter links than 483~km do occur, for example between adjacent planes. There, a free seat could inject into both terminals during acquisition, and the direction check of Section~\ref{sec:freeseats} protects round zero. The \tpaid{two-way seats} that see both directions lie below the orbit along the whole link (Figure~\ref{fig:zones}c).

\subsection{The Hold-Cost Floor}

\para{Hold rate}
On a shared circular orbit, the geometry is rigid in the co-rotating frame, and $\symref{d}{d}$ is a fixed radial displacement. By~\eqref{eq:cw}, holding it against the gravity gradient takes the continuous acceleration
\begin{equation}
    \mathrm{d}\dv/\mathrm{d}t = 3 n^2 d, \qquad 3n^2 \approx 3.6 \times 10^{-6}\ \mathrm{s}^{-2}.
    \label{eq:dv}
\end{equation}
This prices hovering. A floor must also exclude every strategy that lets natural motion do part of the work.

\begin{proposition}[Two-body hold-cost floor]\label{prop:floor}
Let a trajectory stay inside a seat with forced offset $\symref{d}{d}$ for a time $\symref{T}{T}$, and let $\Delta\dot{x}$ and $\Delta y$ be the net changes of its radial velocity and along-track position over the hold. Then its spent $\symref{dv}{\dv}$ satisfies
\begin{equation}
    \dv \ge 3n^2 d\,T - |\Delta\dot{x}| - 2n\,|\Delta y|.
    \label{eq:floor}
\end{equation}
Hovering attains the bound.
\end{proposition}
\begin{IEEEproof}
Integrating the radial equation of~\eqref{eq:cw} over $[0,T]$ gives $\int f_x\,dt = \Delta\dot{x} - 2n\,\Delta y - 3n^2 \int x\,dt$. Every point of the seat lies at least $\symref{d}{d}$ below the reference, so $-3n^2\int x\,dt \ge 3n^2 d\,T$. The thrust magnitude is never below its radial component, so $\dv \ge \int f_x\,dt$, which gives~\eqref{eq:floor}. Hovering has $\Delta\dot{x} = \Delta y = 0$, $f_x = 3n^2 d$, and $f_y = 0$ throughout, and meets the bound with equality.
\end{IEEEproof}

\para{The correction stays constant}
The correction of~\eqref{eq:floor} depends only on how the hold starts and ends, so moving around inside the seat gains only a fixed amount. A craft that stays in the seat moves along the track by at most the seat length $\symref{wy}{w_y}$. Its radial speed changes by at most $2\symref{vb}{v_b}$, where $\symref{vb}{v_b}$ is the largest speed of the craft relative to the seat. The correction is, therefore, at most $2\symref{vb}{v_b} + 2n\,\symref{wy}{w_y}$ and does not grow with $\symref{T}{T}$, while the floor does. We express it as hold time, $\symref{T0}{T_0} = (2\symref{vb}{v_b} + 2n\,\symref{wy}{w_y})/(3n^2 d)$. We call the on-axis seat 1~km from a terminal on the 1000~km link the canonical seat. Each m/s of $\symref{vb}{v_b}$ adds about two hours there, and we take 1~m/s in favor of the attacker. On the line of sight, $\symref{wy}{w_y}$ is the whole chord, but drift along it saves at most the time a free drifter needs to cross it, $\symref{Ts}{T_s} = (8R/3nL)\ln(L/X)$ (Appendix~\ref{app:proofs}). Seats on the line of sight thus have $\symref{T0}{T_0} = 2\symref{vb}{v_b}/(3n^2 d) + \symref{Ts}{T_s}$, about 1.4 days at the canonical seat.

\para{Free motion only visits}
Two force-free families keep a radial offset at no cost, and neither holds a seat. A craft on a slightly lower circular orbit drifts along-track at about $\tfrac{3}{2}nd$ and sweeps through the seat within minutes, which the term $2n\,\Delta y$ in~\eqref{eq:floor} captures. A craft on the drift-free $2{:}1$ ellipse of~\eqref{eq:cw} around a free orbit passes the seat once per orbit, for a few minutes. It misses every round in the rest of the orbit. The protocol demands presence at every round but not between rounds, and rounds at most $\symref{Tmax}{\Tmax}$ apart leave too little time to gain from leaving (Proposition~\ref{prop:parttime}).

\begin{proposition}[Part-time presence saves little]\label{prop:parttime}
Let a trajectory be inside the seat at a sequence of instants spaced at most $\Tmax \ll 1/n$ apart, known to it in advance, and let its speed relative to the seat stay below $\symref{vb}{v_b}$ throughout. Whatever it does in between, over a time $\symref{T}{T}$ it spends at least
\begin{equation}
    \dv \ge 3n^2 d\,\bigl[(1 - c\,(n\Tmax)^2)\,T - T_0\bigr],
    \label{eq:parttime}
\end{equation}
with $c = 3 + \symref{vb}{v_b}/(nd)$.
\end{proposition}

The correction is small. At the canonical seat, it is 0.2 percent at the 10~s spacing of Table~\ref{tab:ref} for $\symref{vb}{v_b} = 1$~m/s, and below seven percent even at one minute. The intuition is simple (Figure~\ref{fig:parttime} in Appendix~\ref{app:figs}). Leaving the seat by $e_k$ between two rounds saves at most $3n^2 e_k \delta_k$ of gravity-gradient work, but leaving and returning within the gap $\delta_k$ costs at least $2e_k/\delta_k$ of thrust. The cost is larger whenever the gap is short against the orbital timescale. Appendix~\ref{app:proofs} gives the proof.

\subsection{Costs and Lifetimes}

\para{The blocking seat}
Table~\ref{tab:price} prices the cheapest seat from which an impostor can block, on the line of sight 1~km from a terminal (\asmbadge{14}). By the symmetry of~\eqref{eq:gap}, it costs the same beside either terminal. On the 1000~km link, a CubeSat-class impostor lasts 4.4 days there, and even a 2~km/s platform lasts under three months. A passive reader of both directions needs only the two-way seat in the read cone, about 0.8 percent cheaper (Table~\ref{tab:cones}), but it gains nothing while ML-KEM holds.

\begin{table}[t]
    \centering
    \caption{The blocking seat, on the line of sight 1~km from a terminal at 550~km, and the days each attacker class can hold it. $\symref{T0}{T_0}$ adds 1.0 to 2.6 days to each lifetime. A passive reader in the two-way seat of the read cone pays 0.5 to 1.6 percent less.}
    \label{tab:price}
    \footnotesize
    \setlength{\tabcolsep}{4pt}
    \begin{tabular}{@{}l r r r r r@{}}
        \toprule
        & & & \multicolumn{3}{c}{Days until the budget runs out} \\
        \cmidrule(l){4-6}
        Link $\symref{L}{L}$ & Offset $\symref{d}{d}$ & m/s per day & CubeSat & Smallsat & Heavy \\
        \midrule
        500~km & 36.0~m & 11.2 & 8.9 & 45 & 178 \\
        1000~km & 72.2~m & 22.5 & 4.4 & 22 & 89 \\
        1500~km & 108~m & 33.7 & 3.0 & 15 & 59 \\
        \bottomrule
    \end{tabular}
\end{table}

\begin{figure}[t]
    \centering
    \includegraphics[width=\columnwidth]{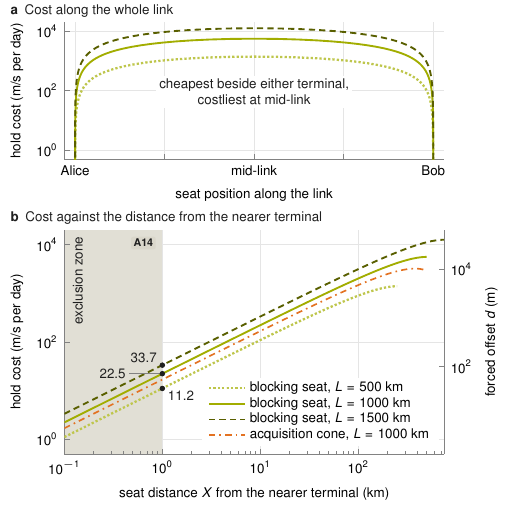}
    \caption{Hold cost of a seat on the line of sight at 550~km altitude. Green lines are blocking seats, light to dark for links of 500, 1000, and 1500~km. (a)~Along the whole link, the cost is symmetric. It is lowest beside either terminal and peaks at mid-link. (b)~The same cost against the distance from the nearer terminal, on log axes. The orange dash-dotted line uses the 17.45~mrad acquisition cone of the inject gate on the 1000~km link. The right axis gives the forced offset $\symref{d}{d}$, to which the cost is proportional. The beige band is the exclusion zone that each terminal keeps clear (\asm{14}), and the dots mark the cheapest seats of Table~\ref{tab:price} at its edge.}
    \label{fig:seatcost}
\end{figure}

\para{Cost along the link}
Near a terminal, the cost grows linearly in $\symref{X}{X}$, since the gap is then close to $XL/2R$, and it reaches thousands of m/s per day near mid-link (Figure~\ref{fig:seatcost}). The cheapest seat for every capability thus sits beside a terminal, which is where the terminal watches.

\section{History-Chained Authentication}\label{sec:protocol}

A single position check can be passed at one fortunate moment, so Alice and Bob keep demonstrating their positions for as long as the link runs, and each demonstration is bound to every earlier one.

\subsection{The Protocol}\label{sec:steps}

\begin{figure}[t]
    \centering
    \includegraphics[width=\columnwidth]{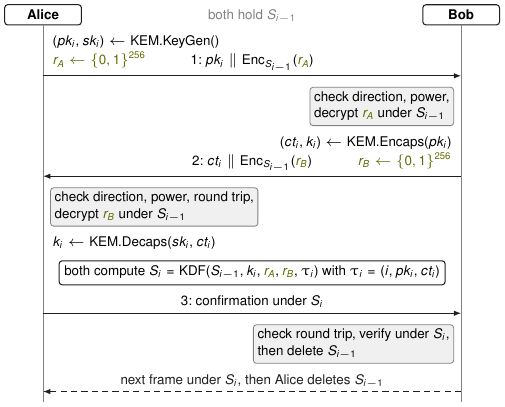}
    \caption{One round $i$ of \pname{} in three messages. Both sides hold $S_{i-1}$. Alice sends a fresh ML-KEM public key and her random bits $\symref{rA}{r_A}$ of round $i$, encrypted with a key from $S_{i-1}$. Bob checks the arrival and encapsulates, which draws a fresh key $\symref{ki}{k_i}$ together with its ciphertext. He returns the ciphertext with his bits $\symref{rB}{r_B}$. Alice confirms $\symref{Si}{S_i}$ in the third message, so both sides measure a round trip. Each side deletes $S_{i-1}$ only after a frame under $\symref{Si}{S_i}$. Green marks the random bits, which only a receiver inside both beams observes. Round zero runs the same exchange without encryption, because no $S_{-1}$ exists, and the fail-closed rule guards it.}
    \label{fig:msc}
\end{figure}

Both endpoints keep a rolling secret $\symref{Si}{\Si{i}}$. Figure~\ref{fig:msc} shows one round.
\begin{enumerate}
    \item \emph{Bootstrap.} Round zero runs the same exchange as every later round, but no $\Si{-1}$ exists. Therefore, it sends $\tpaid{\symref{rA}{r_A}}$ and $\tpaid{\symref{rB}{r_B}}$ in the clear and derives $\Si{0} = \KDF(k_0, \tpaid{r_A}, \tpaid{r_B}, \tau_0)$, the start of the chain. The checks and the fail-closed rule of Section~\ref{sec:checks} guard it. An attacker that already sits on the on-axis seat passes round zero, so security comes only with the paid history that follows.
    \item \emph{Rounds.} At random times, never more than $\symref{Tmax}{\Tmax}$ apart, they run a round of three messages. Alice sends a fresh ML-KEM public key $pk_i$~\cite{fips203}, and Bob encapsulates to it and returns the ciphertext $ct_i$, so both obtain $\symref{ki}{k_i}$. With these two messages, each side also sends the fresh random bits of round $i$, \tpaid{$\symref{rA}{r_A}$} and \tpaid{$\symref{rB}{r_B}$}, under authenticated encryption keyed from $\Si{i-1}$. In the third message, Alice confirms $\symref{Si}{\Si{i}}$ to Bob. Each side thus measures the delay from its own message to the reply.
    \item \emph{Update.} Each side computes
    \[ \Si{i} = \KDF(\Si{i-1}, k_i, \tpaid{r_A}, \tpaid{r_B}, \tau_i) \]
    with the round transcript $\tau_i = (i, pk_i, ct_i)$.
    \item \emph{Checks.} Before adopting $\symref{Si}{\Si{i}}$, each side checks that the peer's light arrives within 50~$\mu$rad of the predicted direction, that received power and round-trip time match the public geometry, and that the levels show one stream. It also checks that the peer holds $\Si{i-1}$, shown by the authenticated decryption of its bits. Keys are separated by direction, round, and attempt. Each ciphertext authenticates the round's public values as associated data, which rejects reflection, replay, and share substitution. The checks do not aim to lock out an active \ac{MitM} on the line of sight. They force it onto the on-axis seat, which it can hold only while its fuel lasts, and security grows with that time.
    \item \emph{Failure and retry.}\label{step:retry} Neither side discards $\Si{i-1}$ until it receives a frame authenticated under $\symref{Si}{\Si{i}}$. A round that fails, through a dropout or a frame that fails authentication, is retried under the last common secret. A retry runs the same round again, and every attempt makes the same demands. Failures stall the chain, so the guarantee stops growing, but they never advance an impostor. The endpoint allows failed attempts only while completed rounds stay at most $\symref{Tmax}{\Tmax}$ apart. Then it stops the link but keeps transmitting, and calls the operator, because failures that persist mean that an attack is starting or ending. Only the operator can resume the chain under the last common secret or start a new round zero, which has no history to check. The chain never runs backwards (Figure~\ref{fig:fsm} in Appendix~\ref{app:figs}).
\end{enumerate}

\para{No unauthenticated entry}
A fresh public key never opens an unauthenticated handshake, because it travels inside a round that must authenticate under $\Si{i-1}$. Every later agreement thus inherits its authentication from the chain, and the ephemeral keys provide forward secrecy and healing, never entry. Only round zero lacks a history. It authenticates nobody, so an active attacker that passes the checks can impersonate either side. The derivation includes the transcript $\symref{taui}{\tau_i}$, so $\symref{Si}{\Si{i}}$ is bound to the exact public values of its round, and the argument does not rely on binding properties of the \ac{KEM}.

\subsection{Two Gates per Round}

The random bits of each round are gated by physics, since only a receiver inside both beams observes both strings, whatever its computing power. The agreed value $\symref{ki}{k_i}$ is gated by mathematics. It never crosses the link, so a passive reader gains nothing from it, even inside both beams and with a leaked secret, unless the hardness assumption falls.

\begin{lemma}[Two-gate composition]\label{lem:gates}
A passive attacker learns $\symref{Si}{\Si{i}}$ only by defeating both gates in the same round. Reading $(\tpaid{r_A}, \tpaid{r_B})$ requires a receiver inside both beams. Obtaining $\symref{ki}{k_i}$ from the public shares requires breaking the \ac{KEM}. Because $\KDF$ takes all inputs and is modeled as a random oracle, missing either leaves $\symref{Si}{\Si{i}}$ unknown, and the chain inherits the stronger of the two gates.
\end{lemma}

\para{Active attackers}
By the lemma, a passive attack is never enough while the KEM holds, so an attacker must turn active and substitute shares. Against an unauthenticated agreement, that would suffice. The chain closes that route, since a substituted round must still authenticate under $\Si{i-1}$, and replacing only Alice's share leaves Bob's reply mismatched at Alice. Therefore, a persistent attack substitutes both directions, from the on-axis seat, over the entire paid history.

\subsection{The \texorpdfstring{$\symref{dv}{\dv}$}{Δv} Floor of an Impostor}

To remain in the chain, the attacker must learn every $\symref{Si}{\Si{i}}$. For that, it needs $\Si{i-1}$, both random strings, and $\symref{ki}{k_i}$. The strings travel in opposite directions, so collecting both means occupying a \tpaid{two-way seat}. The value $\symref{ki}{k_i}$ cannot be read at all, only substituted, which requires the \tpaid{on-axis seat}. This is the role of the KEM. It forces the attacker to be active, on the on-axis seat, at every round, its most expensive case, which Section~\ref{sec:gate} priced. Within that case, we grant it the on-axis seat with the smallest offset $\symref{d}{d}$.

\begin{proposition}[$\symref{dv}{\dv}$ floor for a persistent impostor]\label{prop:bill}
Let the protocol run on a shared circular orbit with rounds at most $\Tmax \ll 1/n$ apart. If one attacking craft (\asm{9}) is accepted as a genuine endpoint at every round over an interval of length $\symref{T}{T}$ and stays on the seat in between, then under \asm{1} to \asm{15} it spends at least
\begin{equation}
    \dv_{\mathrm{attacker}} \ge 3n^2 d\,(T - T_0),
    \label{eq:bill}
\end{equation}
with $\symref{d}{d}$ the forced offset of the on-axis seat. A budget $\symref{dvmax}{\dv_{\max}}$ thus buys at most $T \le \symref{T0}{T_0} + \dv_{\max}/(3n^2 d)$ of sustained impersonation.
\end{proposition}

\emph{Proof idea.} Acceptance at a round requires the on-axis seat at that round, because only there can the attacker suppress Alice's signal. A round the attacker misses either completes between the genuine endpoints, which locks the attacker out, or is retried while the attacker stands unaccepted. Leaving between rounds lets Alice's frames reach Bob, where they fail authentication (step~\ref{step:retry}), and saves at most the fraction $c\,(n\Tmax)^2$ of Proposition~\ref{prop:parttime}. Appendix~\ref{app:proofs} gives the proof.

\para{Whose interval counts}
The interval $\symref{T}{T}$ is the attacker's own participation, from the round it entered the chain, at the bootstrap or with a leaked secret, to the last round it passed.

\para{Several links}\label{sec:severallinks}
Each link authenticates Alice on its own, with its own chain and checks, and her links share no key. They can still combine their security, a step we leave out of scope. Her partners keep their keys (\asm{2}) and can talk to each other securely. For example, Alice derives a combined key pair from all her link secrets, and the partners compare the public key each received. Her lines of sight meet only at her, and each terminal keeps its zone clear (\asm{14}), so one rigid craft sits on one line of sight at most (\asm{15}). A link without a craft heals at its next round, so only a craft on every link learns or replaces the combined key. Against one craft per link, the most that \asm{9} admits, the partners can combine once the most expensive link has outlived the feared budget (Proposition~\ref{prop:bill}).

\subsection{Security Experiment}\label{sec:experiment}

\begin{definition}[Budget-conditioned authentication]\label{def:exp}
Honest endpoints on a shared circular orbit run \pname{} with rounds at most $\Tmax \ll 1/n$ apart, and $\KDF$ is a random oracle. The experiment against an adversary $\mathcal{A}$ runs as follows.
\begin{enumerate}
    \item \emph{Setup.} The challenger starts the round clock. Round zero runs like every later round, so $\mathcal{A}$ may attack it.
    \item \emph{Channel.} $\mathcal{A}$ is the position-gated attacker of Section~\ref{sec:threat}. It sees all public shares. It obtains $(\tpaid{r_A}, \tpaid{r_B})$ at round $i$ only while it occupies a two-way seat at that round's time, and it delivers a message as the peer only from the on-axis seat.
    \item \emph{Reveal.} At any round $j$, $\mathcal{A}$ may query a reveal oracle that returns $\Si{j}$. This models the key leak.
    \item \emph{Budget.} The challenger records the $\symref{dv}{\dv}$ that $\mathcal{A}$ spends.
    \item \emph{Win.} $\mathcal{A}$ wins at round $i$ if a genuine endpoint commits $\symref{Si}{\Si{i}}$ as authentic while no session of its intended peer holds the same $\symref{Si}{\Si{i}}$ for round $i$.
\end{enumerate}
$\mathrm{Adv}^{\mathrm{auth}}_i(\mathcal{A})$ is the probability that $\mathcal{A}$ wins at round $i$.
\end{definition}

\begin{theorem}[Budget-conditioned authentication]\label{thm:exp}
In the experiment of Definition~\ref{def:exp}, let $j$ be the last revealed round, or the bootstrap round $0$ if none. If $\mathcal{A}$ spends less than $3n^2 d\,[(1 - c\,(n\Tmax)^2)(t_i - t_j) - \symref{T0}{T_0}]$ of $\symref{dv}{\dv}$ between $t_j$ and the time $t_i$ of round $i$, then
\begin{equation}
    \mathrm{Adv}^{\mathrm{auth}}_i(\mathcal{A}) \le (i - j + 1)\,\bigl(\varepsilon_{\mathrm{KEM}} + \varepsilon_{\mathrm{AEAD}}\bigr) + q_H / 2^{256},
    \label{eq:composed}
\end{equation}
where $\varepsilon_{\mathrm{KEM}}$ is the IND-CCA advantage against the \ac{KEM}, $\varepsilon_{\mathrm{AEAD}}$ the INT-CTXT advantage against the authenticated encryption, and $q_H$ the number of queries of $\mathcal{A}$ to $\KDF$.
\end{theorem}

\emph{Proof idea.} An adversary below the budget missed the on-axis seat at some round after $j$, which leaves it without the next secret unless it breaks the \ac{KEM} or forges a ciphertext. Appendix~\ref{app:proofs} gives the sketch.

\para{Reading the bound}
Read constructively, the bound states continuity. Each secret requires custody of its predecessor, and each round's fresh entropy requires that round's seat. After an interval $\symref{T}{T}$ of accepted rounds, the partner is the craft on the predicted orbit unless a standardized primitive was broken or an impostor paid the $\symref{dv}{\dv}$ of~\eqref{eq:bill} while escaping surveillance throughout. The bound carries weight only once $t_i - t_j$ exceeds $\symref{T0}{T_0}$, about 1.4 days at the canonical seat. Since a budget caps the spend, the bound covers every adversary whose budget lies below it. The theorem takes the channel rule of Definition~\ref{def:exp} as given, which Sections~\ref{sec:threat}, \ref{sec:gate}, and~\ref{sec:freeseats} argue physically for one rigid craft on the link while Alice transmits (\asm{2}, \asm{9}, \asm{15}).

\subsection{In-Band Checks and the Exclusion Zone}\label{sec:checks}

The chain leaves a surviving attacker on the axis beside a terminal, nearly at rest relative to it, close by, and in a predictable direction. Seven link-only checks expose intruders, and none needs a secret. Angle of arrival and round-trip time expose a source off the predicted direction or range, scan-epoch timing a reply not caused by the terminal's own scan, and the power and level checks a craft that overpowers. A cat's-eye echo exposes optics nearby, and wavefront curvature a source within a few kilometers. Only this check needs new hardware, a second aperture.

\para{What the checks test}
Each terminal runs the checks on its own tracking field of view, whose first kilometer is its exclusion zone. They do not authenticate the far partner, which shows none of these signatures, so a source that shows one is an intruder whatever it transmits. The satellite then calls the operator by radio or by a small unscheduled maneuver, which the tracking of \asm{6} sees and no craft can block. The wavefront check measures distance through the curvature across a 30~cm baseline, though a collimated beam wider than the baseline flattens it. The cat's-eye check probes for nearby optics, which retroreflect a probe~\cite{du2024catseye} and return the echo within microseconds instead of milliseconds. The round-trip check catches only a naive impostor, since one nearer than the partner can add delay.

\para{Round zero}
Round zero rests on these checks. The public catalog places the partner only to about a kilometer, 1~mrad at 1000~km. \pname{}, therefore, adds an ephemeris that the operator signs, sent over any path other than the link and verified under a public key, so the satellite needs no secret. It places the partner to about 10~m, so the direction check holds from round zero on.

\para{Fail-closed rule}
A fail-closed rule adds to these checks at round zero. The terminal finishes its scan of the uncertainty cone before it locks, at the cost of one scan period, and scans only before the first lock, because a scan breaks the lock. If a reply arrives from an inconsistent direction or epoch, or more than one source responds where the ephemeris predicts one, the terminal aborts and retries instead of accepting a wrong partner. An attack on acquisition that the scan sees thus ends in a visible denial of service, which suits a recovery link.

\para{Watching the exclusion zone}
Monitoring orbits shows only the absence of known objects. Each terminal, therefore, watches only its own zone, about 440~m$^3$ in the known direction of the partner. Starlink's nearly 30{,}000 star trackers already detect about 30 million object transits a day~\cite{spacex2026stargaze}. Star trackers see small craft far beyond the zone. The one on BIROS imaged a passive 1U CubeSat from 50~km down to 50~m~\cite{ardaens2018asr}, and those on Starling tracked 6U CubeSats at 50~km~\cite{kruger2026starfox}. At 1~km, the 1U is 2500 times brighter. A nanotube black that passed launch and vacuum tests~\cite{theocharous2014vanta} reflects 0.036 percent at 700~nm~\cite{lehman2018blacks}, so even a coated 1U at 1~km outshines the bare one at 50~km four to nine times, for a bare albedo of 0.1 to 0.2.

\para{No place to hide}
Eclipse gives no cover, since the impostor must stay for days. The Starling star tracker sees stars fainter than magnitude 7~\cite{bct2025nst}. Once per orbit, the sun stands most nearly behind it. It then sees a coated 1U in any attitude out to 6~km on the 1000~km link, and to 3~km in the worst season of the 53$^\circ$ shells. The cat's-eye check sees a craft's optics at any time. Only a craft built for optical stealth passes both, such as one with a mirror front scattering under $3 \cdot 10^{-6}$ per steradian and isolators behind its optics.

\subsection{Ways to Suppress the Genuine Signal}\label{sec:freeseats}

A craft that co-orbits near Alice sits inside both cones that open from Bob, so it reads Bob and injects into Bob at no hold cost. Within about 9~km of her, it lies inside Bob's tracking field of view. It still cannot pass as Alice, because she keeps transmitting toward Bob, and her stream and its own add at his detector. It must suppress her signal, and Table~\ref{tab:ways} in Appendix~\ref{app:figs} lists every way with the check that stops it. Canceling is impossible (\asm{13}), blocking needs the on-axis seat that Propositions~\ref{prop:floor} and~\ref{prop:bill} price, and jamming only denies service.

\para{Steer a beam away}
A four-quadrant sensor follows the centroid of all the light it sees, so the craft can pull Bob's tracking, and with it his beam, toward itself. At 0.14~mrad off Alice, the beam delivers 2 percent of its peak power to her, near the 238~pW down to which the terminal tracked in orbit~\cite{rueddenklau2024acquisition}, and she loses track of him. Therefore, each terminal checks at every round, round zero included, that the light arrives within 50~$\mu$rad of the direction that the signed ephemeris and its star tracker predict. Otherwise, it aborts the round but keeps transmitting along the predicted direction. Its $3\sigma$ knowledge requirements per axis, 24~$\mu$rad of attitude, which two star tracker heads of 1.5~arcsec meet~\cite{nasa2026soagnc}, 20~$\mu$rad of alignment, 10~$\mu$rad of angle reading, 14~$\mu$rad of ephemeris at 1000~km, and 1.1~$\mu$rad for 1~ms of time, give 36~$\mu$rad in root sum square. Under $2 \cdot 10^{-4}$ of honest rounds retry, while a capture must move the beam almost three times as far. Within the threshold plus that error, Alice receives a quarter of the peak and keeps tracking Bob. A craft near Bob that pulls Alice's tracking fails her check in the same way, so even two free craft cannot silence her.

\para{Overpower her at the bit decision}
Added light cannot lower Alice's level (\asm{13}). At a receiver with a fixed threshold, the craft can only turn her zeros into ones, so the merged frame fails authentication. A receiver that adapts its threshold to the mean level, which favors the attacker (\asm{8}), reads the craft's bits where its level exceeds hers, and with a leaked $\Si{i-1}$ its frame passes. A power ceiling cannot stop this, since an override fits under 2~dB once Alice points 35~$\mu$rad off. The level check does. Pointing moves the two levels of one stream only over milliseconds, while the craft's independent bits split them and its light steps them where it starts or stops. The craft cannot see Alice's power dips to hide in them. In the channel simulation of the artifact, the check caught 99.9 percent of second streams at 6 percent of the decoding level, with no alarm in 6000 honest frames. The power check, which guards the linear range of the level check, reads the photocurrent ahead of any gain control, since bright light can blind a detector~\cite{lydersen2010hacking}, and rejects a round 2~dB above the level of perfect pointing, which pointing errors never exceed.

\para{Acquisition}
The direction check also guards the scan of round zero. A free seat that passes it lies within 1.2~km of Alice, where the level check exposes it while she transmits (\asm{2}), so a free seat buys only a denial of service. To silence Alice, the impostor must block her beam from the paid on-axis seat. While she is silent for her own reasons, as in a reboot, a free craft within 1.2~km passes Bob's checks. With the current secret or at round zero, it is accepted until she transmits again and the split chains call the operator.

\section{Beyond the Shared Orbit}\label{sec:general}

Leaving the shared orbit does not lower the attacker's cost at equal chord length. We support this with numerical arguments. On a shared orbit, the link is stationary in the co-rotating frame, and only the gravity-gradient hold remains. Every other geometry adds relative motion between the attacker's best free orbit and the seat, which raises the cost, shortens the dwell, or both.

\para{Intermediate orbits give isolated contacts}
An attacker on a free circular orbit between two shells aligns with the sweeping line of sight only at isolated instants, since three craft on distinct circular orbits lie on one line only when a sum of sine waves with different frequencies is zero (Figure~\ref{fig:rider} in Appendix~\ref{app:figs}). On coplanar 550 and 1150~km shells, this gives contacts of 4.1 to 7.6~minutes inside the read cone, at most once per 13.6-hour synodic cycle, too rare to stay in the chain. A contact inside the acquisition cone can last the whole 49-minute window in which the pair is in range, which buys at most round zero. Partners on different orbits keep the current secret between their contact windows. The bounds of Section~\ref{sec:protocol} assume a link that stays in sight, and the evaluation uses only such links.

\para{Tightness}
Between adjacent planes of equal radius, the chord turns. A chord-riding seat at a fixed fraction of it pays the floor of its own depth to about $10^{-5}$ at the 1~km seat, while a rigid standoff pays up to about twice the floor for planes $5^\circ$ apart with a shortest chord of 1000~km.

\para{Scaling}
At a fixed seat, the floor grows about linearly with $\symref{L}{L}$ and falls as $R^{-4}$, about 1400 times from 550~km to geostationary orbit. A 100~km link costs 2.2~m/s per day at the 1~km seat, so a link that decides a recovery should be longer than about 500~km. Higher shells fly longer links, and a 3000~km link at 2000~km altitude costs within ten percent of a 1500~km link at 550~km.

\section{Evaluation}\label{sec:eval}

We ask four research questions.
\textbf{\rqref{1}} What does a persistent attack cost on the crosslinks of a real constellation?
\textbf{\rqref{2}} How long does each attacker class last, and does electric propulsion change that?
\textbf{\rqref{3}} Does the analytic floor survive real dynamics and real ephemerides?
\textbf{\rqref{4}} What does the defender pay?

\subsection{Setup}

\para{Data}
We downloaded the public CelesTrak element sets of all 11{,}109 Starlink satellites in the catalog on 26 October 2026 and propagated them with SGP4 from their median epoch. Removing satellites that are raising, decaying, or drifting leaves 9868 satellites in eight shells between 356 and 572~km altitude.

\para{Topology}
SpaceX does not publish its laser topology. We assume the +Grid pattern common in constellation networking research, in which each satellite links to its two in-plane neighbors and to one satellite in each adjacent plane~\cite{handley2018delay, bhattacherjee2019topology}. Of these links, 17{,}433 keep their line of sight above 80~km altitude for a full orbit.

\para{Metric}
For each link, we price the blocking seat 1~km from a terminal, on the instantaneous SGP4 chord, and average over one orbit. In-plane neighbors fly nearly one orbit, where this cost is the proven floor (\rqref{3}). Between planes, it is a lower estimate without a proven floor (Section~\ref{sec:general}). The chord-riding seat, the cheapest attack we know, sits farther out on longer chords and pays a median 35 percent more than this estimate. Lifetimes in \rqref{1} include $\symref{T0}{T_0}$ and the factor of Proposition~\ref{prop:parttime}, except in Table~\ref{tab:starlink} and Figure~\ref{fig:starlink}.

\para{Parameters and artifact}
We keep the reference terminal of Table~\ref{tab:ref}, because SpaceX does not publish its terminal parameters. Starlink's terminals reach 5400~km~\cite{ali2026leotopology}, so they have narrower cones, and in cost, the reference terminal favors the attacker. A reproducible artifact recomputes every bound, table, and propagation, and a harness checks 182 printed values against them.

\subsection{RQ1: Cost on the Real Starlink Constellation}\label{sec:rq1}

\para{Links}
In-plane links have a median length of 1386~km and cost a median 32.7~m/s per day (Table~\ref{tab:starlink}). Inter-plane links are shorter, a median 321~km, because adjacent planes of the dense shells lie only $2^\circ$ to $4^\circ$ apart in node, and cost a median 7.5~m/s per day.

\begin{figure}[t]
    \centering
    \includegraphics[width=\columnwidth]{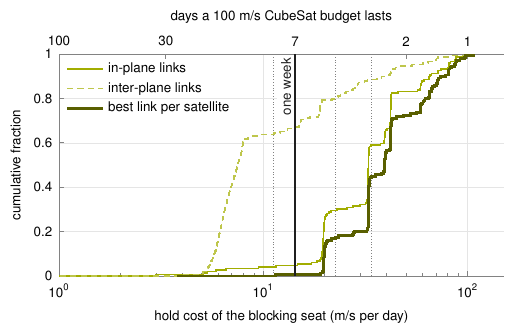}
    \caption{Cost of the blocking seat on 17{,}433 +Grid crosslinks of the Starlink constellation, from public element sets of 26 October 2026 propagated with SGP4. The thick curve takes, for each of the 9868 satellites, its most expensive link, the cost a recovering satellite can force. The top axis gives the lifetime of a 100~m/s budget before $\symref{T0}{T_0}$, and the solid guide marks one week. Dotted lines mark the shared-orbit values of Table~\ref{tab:price}.}
    \label{fig:starlink}
\end{figure}

\para{The best link decides}
Each link authenticates on its own (Section~\ref{sec:severallinks}), so the most expensive link of a satellite sets how soon the constellation can trust it again. The most expensive +Grid link of each satellite is a median 1684~km long and an in-plane link for 91 percent of satellites. It costs a median 39.3~m/s per day (Figure~\ref{fig:starlink}). For 98.7 percent of satellites, that link exhausts a CubeSat-class craft within a week, for 98.9 percent a smallsat within 30 days, and for 99.4 percent a heavy platform within six months. Restricted to the 1500~km range of the reference terminal, the CubeSat share is 84.5 percent of the satellites with such a link.

\para{Any partner in the shell}
We searched each shell for the most expensive partner whose line of sight stays clear and in range for a full orbit (Table~\ref{tab:starlink}). At the 5400~km reach of Starlink's terminals, every satellite finds a partner whose link averages at least 2770~km and costs at least 69.6~m/s per day, so a CubeSat-class craft runs dry in about two days. Through an 80~mm aperture, the reference beam arrives at 5400~km as strongly as at 1350~km, so the reference checks apply. Within the 1500~km range of the reference terminal, every satellite of the two dense shells, which hold 6782 of the 9868 satellites, finds a partner at least 1000~km away on average, which exhausts a CubeSat-class craft within 5.7 days. Over all shells, the best partner in range does so within a week for 96.4 percent of those with one.

\begin{table}[t]
    \centering
    \caption{Blocking-seat cost on the real Starlink geometry, with the lifetime of a 100~m/s budget at the median cost, before $\symref{T0}{T_0}$.}
    \label{tab:starlink}
    \footnotesize
    \setlength{\tabcolsep}{4pt}
    \begin{tabular}{@{}lrrrrr@{}}
        \toprule
        & & Median & \multicolumn{2}{c}{m/s per day} & CubeSat \\
        \cmidrule(lr){4-5}
        Population & Count & $\symref{L}{L}$ (km) & 5th pct. & Median & (days) \\
        \midrule
        In-plane links & 9549 & 1386 & 12.9 & 32.7 & 3.1 \\
        Inter-plane links & 7884 & 321 & 5.5 & 7.5 & 13.4 \\
        Best link per satellite & 9868 & 1684 & 19.7 & 39.3 & 2.5 \\
        Best partner in 1500~km & 9168 & 1390 & 21.3 & 32.7 & 3.1 \\
        Best partner in 5400~km & 9868 & 4449 & 85.4 & 104.8 & 1.0 \\
        \bottomrule
    \end{tabular}
\end{table}

\para{Robustness}
A satellite with fewer than the four lasers of +Grid lacks some inter-plane links, but any layout keeps the in-plane neighbors, and counting in-plane links alone, which rest on the proven floor, lowers the shares by up to 1.2 percentage points. Most satellites in our snapshot fly at 463 to 483~km, after SpaceX began lowering its 550~km shell~\cite{walter2026starlinklower}, which raises the cost per kilometer of link by about five percent.

\callout{Answer to \rqref{1}.}{Against one CubeSat-class craft per link, almost every satellite of a real constellation has a crosslink that exhausts its craft within a week, and within the reach of Starlink's own terminals, every satellite has a partner whose link does so in about two days.}

\subsection{RQ2: Attacker Lifetimes}\label{sec:rq2}

\para{Lifetimes}
On the 1000~km link, a CubeSat lasts 4.4 days, a smallsat 22 days, and a heavy platform 89 days (Table~\ref{tab:price}). The chain cannot stop a shorter attack, but ends every attack within this time plus $\symref{T0}{T_0}$, in a lockout or in exposure. The operator trusts a key once the link has run that long past it without a stop or a call. A year of deception demands 4 to 12~km/s on the three links.

\para{Electric propulsion}
Electric propulsion offers more $\symref{dv}{\dv}$ per kilogram but must also supply a mean acceleration of $2.6 \times 10^{-4}$~m/s$^2$ at the canonical seat, continuously and through every eclipse. Gridded ion engines on craft below about a tonne and Starlink-class Hall thrusters on the lightest Starlink craft reach it (Figure~\ref{fig:propulsion} in Appendix~\ref{app:figs}). With Hall thrusters, a year on the 1000~km link still burns 28 percent of the craft's mass, as do four months on a partner link.

\callout{Answer to \rqref{2}.}{Each budget class runs dry within three months on a 1000~km link, and a year of impersonation there needs about 8~km/s, close to the record 11.5~km/s that Dawn delivered~\cite{garner2018dawn}. Electric propulsion enlarges the budget, and the daily cost stays the same.}

\subsection{RQ3: Cross-Validation of the Floor}\label{sec:rq3}

\para{Dynamics}
A numerical inverse-dynamics propagation computes the thrust that keeps a craft on the canonical seat under three force models. On a 550~km equatorial orbit, the two-body cost reproduces $3n^2 d\,T$ to a ratio of 1.00001, Earth's oblateness ($\symref{J2}{J_2}$) raises it by 0.23 percent, and drag adds under $10^{-4}$. At inclination $i$, the orbit-averaged ratio follows $1 + 2.5J_2(R_\oplus/R)^2(1 - 1.5\sin^2 i)$ and drops below one above $54.7^\circ$ (Figure~\ref{fig:j2} in Appendix~\ref{app:figs}). At the $53^\circ$ of most Starlink shells, $\symref{J2}{J_2}$ adds only 0.01 percent, and on polar orbits the hold costs about 0.1 percent less than the floor, which moves no reported figure.

\para{Real ephemerides}
Chord and radius vary along the SGP4 trajectories. On nearly every link, the floor at the mean geometry matches its orbit average to within 0.1 percent between in-plane neighbors and 0.04 percent between planes. The closed form also predicts the peak rotation of the line of sight, a median 1.4~mrad/s between planes, to a median ratio of 1.00. NASA's Shuttle handbook budgets station keeping directly below or above a target within about 3 percent of the same law~\cite{nasa2005gpo}.

\callout{Answer to \rqref{3}.}{The analytic floor matches numerical propagation with $\symref{J2}{J_2}$ and drag to within 0.25 percent at every inclination, and its average along SGP4 orbits to within 0.1 percent. The propagated cost lies above the floor, except above $54.7^\circ$ of inclination, where it lies up to about 0.1 percent below.}

\subsection{RQ4: Cost for the Defender}\label{sec:rq4}

\para{Bandwidth and state}
A round with ML-KEM-768 carries a 1184-byte public key, a 1088-byte ciphertext, 64 random bytes, and about 150 bytes of check, confirmation, and authentication fields in three messages, about 2.5~KB in total. At the fastest cadence of one round per second, this is 0.02 percent of a 100~Mbps link. The secret state is two 32-byte values, and the in-band checks reuse measurements the terminal already makes.

\para{Hardware}
Only the wavefront check needs new hardware, a second aperture about 30~cm from the first. The direction check needs two star tracker heads of 1.5~arcsec, an alignment of star tracker and terminal held to 20~$\mu$rad, and an angle reading to 10~$\mu$rad (Section~\ref{sec:freeseats}). It must stay below the beam's divergence, so narrower beams need a finer check. These requirements, the false-alarm rates on real receivers, and how often outages call the operator need tests on real terminals. A sensor less accurate than required raises false alarms, not false acceptances.

\callout{Answer to \rqref{4}.}{\pname{} occupies at most 0.02 percent of the link, and on a satellite with two star tracker heads its baseline needs no new hardware.}

\section{Discussion and Limitations}\label{sec:discussion}

\para{Deployment}
We recommend running the chain continuously rather than switching it on after a key leak. A leak, noticed or not, then needs no new round zero. The attacker that holds the leaked secret must take the on-axis seat at once, and the guarantee grows again from that round (Section~\ref{sec:overview}). A satellite needs no crosslink key before launch, because the chain authenticates from scratch in orbit. A key that leaks during assembly thus never enters it.

\para{Limitations}
We state five limits.
\begin{enumerate}
\item[\limdef{1}] \emph{Collusion.} No coalition avoids the line of sight. Free craft beside it cannot silence Alice (Section~\ref{sec:freeseats}) and learn no key while ML-KEM holds, so one craft must block at every round. A craft relieving a dry one adds only its budget and one $\symref{T0}{T_0}$. Two craft on one link can also take turns, which the floor does not price. Sliding from 1 to 4~km on the 1000~km link and flying back, such a pair still pays about 3.9~m/s per day, a sixth of the floor. The gain shrinks with link length, to a factor of about 1.7 at 2770~km and to none at 4449~km, the shortest and the median partner link of \rqref{1}. \asmbadge{9} admits one craft per link, like position verification~\cite{chandran2009position, buhrman2014position}.
\item[\limdef{2}] \emph{The exclusion zone.} \asmbadge{14} sets the standoff, a parameter of the bound. Near a terminal, the cost is proportional to it, 22.5~m/s per day per kilometer on the 1000~km link, so a 100~m standoff stretches the 4.4 days to about 44. Section~\ref{sec:checks} makes a strong case for 1~km, pending a flight test. Wiretap secrecy in space also assumes such zones, with radii up to 16~km~\cite{vazquezcastro2021qkd}. An operator that fears a craft built for optical stealth prices the link at the standoff it can enforce.
\item[\limdef{3}] \emph{Operator and replacement.} The primitive authenticates only the spacecraft and its position, not the operator or its ground segment. Any implant runs on the true craft in its true orbit, so \asm{2} draws the line at a leak of key material. An attacker that removes Alice and flies her trajectory passes every check, but destroying her leaves trackable debris.
\item[\limdef{4}] \emph{The beam model.} The read cone depends on the Gaussian beam of \asm{5}, but the forced offset falls only linearly in its width. A cone ten times wider removes about 15 percent of the cost on the 500~km link. Authentication does not depend on \asm{5}, since the on-axis seat has no cone slack.
\item[\limdef{5}] \emph{Tethers and booms.} A module on a tether below a bus on a free orbit can hold the blocking seat with almost no propellant, because the tether carries the load. Parts that reach a distance $s$ from the center of mass lower the paid offset to $d - s$, only a few percent for a craft a few meters across, and \asmbadge{15} excludes longer structures. A tether spans at least $\symref{d}{d}$, 72~m at the canonical seat and 72~m more per kilometer of standoff, and must damp its swing, with a period of 55 minutes at 550~km, to the sub-meter precision of the seat, which no flown tether has shown. Its bus sits about $L/2R$ above the beam, $4^\circ$ on the 1000~km link, outside the exclusion zone. Against a tether, the defense is a sensor that watches a few degrees above the beam, rather than the fuel floor.
\end{enumerate}

\section{Related Work}\label{sec:related}

Apart from the poster abstract~\cite{birkel2026poster}, no prior approach both recovers after total key compromise and prices the attacker's effort over time (Table~\ref{tab:related}).

\begin{table}[t]
    \centering
    \caption{Comparison with the closest approaches. \yes{} yes, \partly{} partly, \no{} no. Ratchets recover only once the attacker turns passive and authenticate only from an authenticated start. Statistical schemes accept at a tuned false-alarm rate and need reference data or extra receivers. Distance bounding usually needs a shared key, and position verification needs several verifiers. \pname{} authenticates from round one, guards round zero by the fail-closed rule, and its bound is worst case only under \asm{9}, \asm{14}, and \asm{15}.}
    \label{tab:related}
    \footnotesize
    \setlength{\tabcolsep}{2.2pt}
    \begin{tabular}{@{}>{\raggedright\arraybackslash}p{3.45cm} c c c c c@{}}
        \toprule
        Approach & \shortstack{Keyless\\recovery} & \shortstack{Authen-\\ticates} & \shortstack{Worst\\case} & \shortstack{Partner\\alone} & \shortstack{Prices\\effort} \\
        \midrule
        Ratchets, post-compromise security~\cite{perrin2016doubleratchet, cohngordon2016pcs} & \partly & \partly & \yes & \yes & \no \\
        Quantum key distribution~\cite{scarani_security_2009} & \no & \no & \yes & \yes & \no \\
        Wiretap and exclusion-zone secrecy~\cite{pan2021geometrical, vazquez-castro_quantum_2024} & \yes & \no & \partly & \yes & \no \\
        Statistical authentication~\cite{topal_physical_2022, abdelsalam_ml-driven_2025, kohli_emma_2026, smailes2023satiq} & \yes & \partly & \no & \partly & \partly \\
        Orbit and track checks~\cite{jedermann2021orbit, lee2026challenge, schaefer2015track, schaefer2016doppler} & \yes & \partly & \no & \no & \no \\
        Distance bounding, position-based crypto~\cite{brands1993distance, capkun2006secure, chandran2009position} & \partly & \yes & \yes & \partly & \no \\
        \midrule
        \pname{} (this work) & \yes & \yes & \partly & \yes & \yes \\
        \bottomrule
    \end{tabular}
\end{table}

\para{Post-compromise security}
The double ratchet~\cite{perrin2016doubleratchet} heals against an attacker that turns passive, since every round folds in a fresh key agreement~\cite{cohngordon2016pcs, cheval2026doubleratchet}. It assumes an authenticated start, and it never heals against an attacker that keeps substituting shares, because substitution is free on a network. \pname{} heals for the same reason. On an \ac{OISL}, however, substitution requires the on-axis seat, so an active attacker pays $\symref{dv}{\dv}$ for every second it stays.

\para{Wiretap secrecy for optical links}
Wyner's wiretap model~\cite{wyner1975wiretap} makes confidentiality attainable whenever the legitimate channel outperforms the eavesdropper's, and optical schemes carry it to satellite links~\cite{endo2015wiretap, pan2021geometrical, vazquezcastro2021qkd, vazquez-castro_quantum_2024}. Their eavesdropper is passive, its share of the light is bounded, and none of them authenticates the sender. V\'azquez-Castro et~al.~\cite{vazquezcastro2021qkd} keep a 2~m telescope 16~km from the satellite and argue that celestial mechanics lets no craft park in the line of sight. Pan and Djordjevic~\cite{pan2021geometrical} exclude the eavesdropper from the line of sight and never place it between the terminals. Our attacker is stronger. It may act, its aperture is unbounded, and it is kept only from the first kilometer of a thin cone. It may even hold the line of sight, at the thrust that we price. Wiretap secrecy, in turn, needs no hardness assumption.

\para{Kinematic and keyless authentication}
Topal et~al. and Farrea et~al. derive key material from orbital motion~\cite{topal_securing_2021, farrea_zero_2025} but remain key-based. Topal and Karabulut Kurt~\cite{topal_physical_2022}, Abdelsalam et~al.~\cite{abdelsalam_ml-driven_2025}, and Kohli et~al.~\cite{kohli_emma_2026} authenticate statistically from Doppler or learned features. Jedermann et~al.~\cite{jedermann2021orbit} and Lee et~al.~\cite{lee2026challenge} check a satellite's signals against its expected orbit from the ground, and Smailes et~al.~\cite{smailes2023satiq} fingerprint radio hardware. Their guarantees are statistical, and none bounds the spoofing effort. The closest terrestrial analogue of our continuity argument checks that an aircraft reports a track~\cite{schaefer2015track} and a motion~\cite{schaefer2016doppler} that stay physically possible over time.

\para{Position-based cryptography and distance bounding}
Brands and Chaum~\cite{brands1993distance} bound a prover's distance by timing against the speed of light, and \v{C}apkun and Hubaux~\cite{capkun2006secure} combine such bounds from several verifiers into verifiable multilateration. Chandran et~al.~\cite{chandran2009position} formalized position-based cryptography and proved position verification impossible against colluding adversaries, a result that extends to quantum colluders who share enough entanglement~\cite{buhrman2014position}. Mauw et~al.\ verify distance-bounding protocols~\cite{mauw2018distancebounding} and formalize collusion with a dishonest prover~\cite{mauw2019postcollusion}, while our prover is honest (\asm{2}). None of these results attaches a cost to occupying a location, and we price the location itself. Tippenhauer et~al.~\cite{tippenhauer2011gps} derived the transmitter placements a \ac{GPS} spoofer needs, and we derive the analogous placements for \ac{OISL} attacks together with their orbital cost.

\para{Satellite security and maneuvering}
Surveys show that satellite security still rests on cryptography~\cite{salim2024, tedeschi2022, bala2025security}, and Abdelsalam et~al.\ name \acp{ISL} an open physical-layer problem~\cite{abdelsalam_physical_2025}. Pricing an adversary's reach in $\symref{dv}{\dv}$ is established in orbital pursuit-evasion~\cite{xu2025delta, venigalla2021deltav}, and rendezvous maneuvers show up in public catalog data, though only after the fact~\cite{serrano2024monitoring}. We adopt only the cost measure and tie it to a security guarantee.

\section{Conclusion}\label{sec:conclusion}

A satellite whose keys have all leaked can still be authenticated by its orbit. An impostor on a laser crosslink must block the genuine beam from a seat below free fall. Holding that seat costs a proven floor of $\symref{dv}{\dv}$, and \pname{} turns that floor into a cost that exhausts any fixed budget at a computable time. On the real Starlink constellation, every satellite has a partner whose link exhausts a CubeSat-class craft in about two days against the best attack we know, at under 0.02 percent of the link. These results rest on at most one rigid attacking craft per link and an enforced exclusion zone. A step that combines several links and switches among partners, a detection bound for the zone, a mechanized model of the chain, and a flight test of the checks are the next steps. After an interval of accepted rounds, the partner is the craft on its predicted orbit unless a standardized primitive fell or an impostor paid the fuel cost unseen. Even with a fallen primitive, an impostor pays fuel for as long as it stays.

\makeatletter\def\UrlBigBreaks{\do@url@hyp}\g@addto@macro\UrlOrds{\do\:}\makeatother
\bibliographystyle{IEEEtran}
\bibliography{bibliography}

\appendices

\section{Proofs}\label{app:proofs}

\begin{IEEEproof}[Proof of Proposition~\ref{prop:parttime}]
Let $e_k$ be the largest radial distance from the seat that the trajectory reaches between instants $k$ and $k+1$, a gap of length $\delta_k \le \Tmax$. Along-track and cross-track excursions leave the tidal term unchanged, so only the radial one matters. Integrating the radial equation of~\eqref{eq:cw} over the whole interval, as in Proposition~\ref{prop:floor}, gives $\dv \ge 3n^2 d\,T - 3n^2 \sum_k e_k \delta_k - 2\symref{vb}{v_b} - 2n\,\symref{wy}{w_y}$, because an excursion reduces the tidal term by at most $3n^2 e_k \delta_k$ per gap.

An excursion also costs thrust. Leaving the seat by $e_k$ and returning within $\delta_k$ forces the radial velocity to reach $e_k/\delta_k$ on the way out and $-e_k/\delta_k$ on the way back, a total variation of at least $2e_k/\delta_k$. The gravity gradient and the Coriolis term of~\eqref{eq:cw} supply at most $(3n^2(d + e_k) + 2n \symref{vb}{v_b})\,\delta_k$ of it. Thrust supplies the rest, so $\dv \ge \sum_k [\,2e_k/\delta_k - (3n^2(d + e_k) + 2n \symref{vb}{v_b})\,\delta_k\,]$.

Weighting the two bounds by $1 - \beta$ and $\beta$ with $\beta = \tfrac{3}{2}(n\Tmax)^2$ cancels every $e_k$, up to a nonnegative remainder that is dropped, and leaves~\eqref{eq:parttime}. The free crossings of Section~\ref{sec:gate} are the case of gaps near an orbital period. There, the correction is of order one and the bound carries nothing, which is why the cadence must stay well below the orbital timescale.
\end{IEEEproof}

\begin{IEEEproof}[Proof of Proposition~\ref{prop:bill}]
To be accepted at a round, the attacker must occupy the on-axis seat at that round's time. A round the attacker misses either completes between the genuine endpoints or fails. If it completes, the attacker is locked out, since it can neither form the next secret nor catch up. If it fails, it is retried as the same round under the last common secret, which demands the seat again while the attacker stands unaccepted. Therefore, acceptance at every round over $[t, t+T]$ implies occupation of the seat at every round. An attacker that stays on the seat in between holds it throughout, so Proposition~\ref{prop:floor} bounds its cost and $\symref{T0}{T_0}$ absorbs the correction.

The on-axis seat spans the chord. At depth $g = y(L-y)/(2R)$ and speed $v$ along it, the radial thrust is $|3n^2 g - 2nv|$, up to changes of radial speed that $2\symref{vb}{v_b}$ covers. Per meter, forward motion thus saves at most $2/(3ng)$ of hold time, at $v = 3ng/2$, and backward motion loses as much since $g \ge d$, so drift saves at most $\int_X^{L-X} 2\,dy/(3ng) \le \symref{Ts}{T_s}$.
\end{IEEEproof}

\begin{IEEEproof}[Proof sketch of Theorem~\ref{thm:exp}]
By Propositions~\ref{prop:parttime} and~\ref{prop:bill}, an adversary below the budget was absent from the seat at some round $m$ with $j < m \le i$. At that round, it either lacked $(\tpaid{r_A}, \tpaid{r_B})$, so $\Si{m}$ is uniform to it under the random oracle, or read them from a two-way seat and lacked $k_m$, which it recovers from the public shares only as an IND-CCA distinguisher against the \ac{KEM}. This lockout is the healing argument of post-compromise security~\cite{cohngordon2016pcs}, with the random bits as a second, physical source of fresh secrets.

An endpoint commits $\symref{Si}{\Si{i}}$ only if all inputs check out. Passing the freshness check at any round without its predecessor requires a valid ciphertext under a key derived from that predecessor, an INT-CTXT forgery. The remaining path, already holding the predecessor, recurses one round back. The induction ends at round $j$, the last secret the reveal oracle handed out, or at round zero, which counts as revealed because $\mathcal{A}$ may have attacked it. A \ac{MitM} keeps one chain with each endpoint and is accepted at a round only from the axis, so the same cases cover it. Summing the per-round terms over at most $i - j + 1$ rounds gives~\eqref{eq:composed}, and $q_H/2^{256}$ bounds the chance that a query guesses a missing input.
\end{IEEEproof}

\section{Notation and Glossary}\label{app:notation}

Table~\ref{tab:notation} lists the symbols and defined terms, one row per concept. In the PDF, each symbol links to its entry.

\begin{table}[t]
    \centering
    \caption{Symbols and defined terms.}
    \label{tab:notation}
    \footnotesize
    \setlength{\tabcolsep}{3pt}
    \begin{tabular}{@{}l >{\raggedright\arraybackslash}p{6.2cm}@{}}
        \toprule
        \symdef{dv}{$\dv$} & Velocity change by propulsion, measure of propellant \\
        \symdef{R}{$R$}, \symdef{n}{$n$} & Orbit radius, mean motion $\sqrt{\mu/R^3}$ \\
        \symdef{mu}{$\mu$}, \symdef{Rearth}{$R_\oplus$}, \symdef{J2}{$J_2$} & Earth's $GM$, equatorial radius, oblateness term \\
        \symdef{x}{$x$}, \symdef{y}{$y$} & Radial and along-track offset from point P \\
        \symdef{L}{$L$}, \symdef{X}{$X$} & Link length, seat distance from the nearer terminal \\
        \symdef{theta0}{$\theta_0$}, \symdef{theta}{$\theta$} & Divergence half-angle of the beam, half-angle of a cone \\
        \symdef{gnat}{$\gnat$} & Gap between chord and orbit at distance $\symref{X}{X}$ \\
        \symdef{d}{$d$} & Forced offset of a seat below free fall, $\gnat - \theta X$ \\
        \symdef{holdrate}{$3n^2 d$} & Hold rate or hold cost, the $\symref{dv}{\dv}$ per unit time to keep a seat \\
        \symdef{T}{$T$} & Interval an impostor holds a seat or is accepted \\
        \symdef{T0}{$T_0$} & Constant correction of the floor, written as hold time \\
        \symdef{Ts}{$T_s$} & Time a free drifter needs to cross the link \\
        \symdef{vb}{$v_b$} & Largest speed of the attacker relative to its seat \\
        \symdef{wy}{$w_y$} & Length of a seat along the track \\
        \symdef{Tmax}{$\Tmax$} & Largest spacing of two completed rounds \\
        \symdef{Si}{$\Si{i}$}, \symdef{ki}{$k_i$} & Chain secret and ML-KEM key of round $i$ \\
        \symdef{rA}{$\tpaid{r_A}$}, \symdef{rB}{$\tpaid{r_B}$} & Fresh random bits of Alice and Bob in round $i$ \\
        \symdef{taui}{$\tau_i$} & Transcript $(i, pk_i, ct_i)$ of round $i$ \\
        \symdef{dvmax}{$\dv_{\max}$} & Attacker budget \\
        \midrule
        Seat & Position from which the attacker uses a capability \\
        Reference point P & Point of the orbit directly above a seat, origin of the frame \\
        \tread{Read cone} & Where a receiver demodulates a beam, 0.59~mrad \\
        Secrecy cone & Where a receiver learns $10^{-9}$ bits per bit, 0.71~mrad \\
        \tinject{Inject cone} & A field of view, 0.65 or 17.45~mrad in acquisition \\
        \tpaid{Two-way seat} & Inside cones from both terminals, sees both ways \\
        \tpaid{On-axis seat} & On the line of sight, needed to block or modify, also called the \tpaid{blocking seat} \\
        Free seat & On a free orbit near one terminal, inside cones from the other only \\
        Exclusion zone & First kilometer of a tracking field of view (\asm{14}) \\
        1~km standoff & Edge of the exclusion zone, the cheapest seat \\
        Canonical seat & On-axis seat at the 1~km standoff, 1000~km link \\
        Chain, round & Secrets $\Si{0}, \Si{1}, \dots$, one exchange that advances them \\
        \bottomrule
    \end{tabular}
\end{table}

\section{Supplementary Figures and Tables}\label{app:figs}

\begin{figure}[t]
    \centering
    \includegraphics[width=\columnwidth]{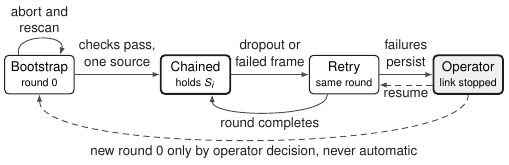}
    \caption{States of one endpoint of \pname{}. A failed round, from a dropout or a frame that fails authentication, is retried as the same round under the last common secret. Failures that persist mark an attack that starts or ends. The endpoint then stops the link and calls the operator, a human in the loop. It never returns to round zero on its own, and only the operator may resume the chain or start a new one. The chain never runs backwards.}
    \label{fig:fsm}
\end{figure}

\begin{figure}[t]
    \centering
    \includegraphics[width=\columnwidth]{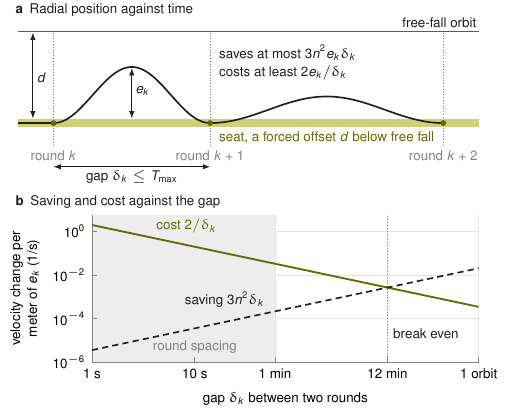}
    \caption{Why part-time presence saves little (Proposition~\ref{prop:parttime}). (a)~The green seat lies a forced offset $\symref{d}{d}$ below free fall, and a trajectory must be inside it at every round, at most $\symref{Tmax}{\Tmax}$ apart. Rising toward free fall by $e_k$ between two rounds saves at most $3n^2 e_k \delta_k$ of gravity-gradient work, but leaving and returning within the gap $\delta_k$ costs at least $2e_k/\delta_k$ of thrust. (b)~Both terms per meter of $e_k$ against the gap at 550~km. The saving grows with the gap and the cost falls, and they break even only at about 12 minutes. At round spacings of seconds to a minute, shaded, the cost exceeds the saving at least 150 times.}
    \label{fig:parttime}
\end{figure}

\begin{figure}[t]
    \centering
    \includegraphics[width=\columnwidth]{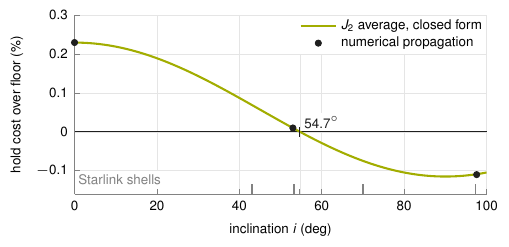}
    \caption{Hold cost of the on-axis seat under Earth's oblateness ($\symref{J2}{J_2}$) against inclination, relative to the analytic floor, at 550~km. The green curve is the orbit average $2.5J_2(R_\oplus/R)^2(1 - 1.5\sin^2 i)$, and the dots are the numerical propagations of Section~\ref{sec:eval}. The floor is undercut only above $54.7^\circ$, and by at most about 0.12 percent. Gray ticks mark the inclinations of the Starlink shells of Section~\ref{sec:eval}.}
    \label{fig:j2}
\end{figure}

\para{Further figures}
Figures~\ref{fig:fsm}, \ref{fig:parttime}, and~\ref{fig:j2} show step \ref{step:retry} of the protocol, Proposition~\ref{prop:parttime}, and \rqref{3} at all inclinations.

\para{Checks against a free craft}
Table~\ref{tab:ways} lists the ways of Section~\ref{sec:freeseats} on the 1000~km link, for a detector noise known to 5 percent. The angle of arrival, mirror angle plus sensor reading, contains the loop residual of 71~$\mu$rad~\cite{rueddenklau2024acquisition}, and the point-ahead angle of 7.3~$\mu$rad applies only to the transmit beam. A single 5~arcsec head across the line of sight would add its roll error of up to 55~arcsec~\cite{nasa2026soagnc}.

\begin{table}[t]
    \centering
    \caption{Ways for a craft near Alice to suppress her signal at Bob, its seat, and the check that stops it, with analytic false-alarm rates.}
    \label{tab:ways}
    \footnotesize
    \setlength{\tabcolsep}{2pt}
    \begin{tabular*}{\columnwidth}{@{\extracolsep{\fill}}l l l l l@{}}
        \toprule
        Way & From her, km & Check & Threshold & False alarm \\
        \midrule
        Cancel & off axis & impossible & & \\
        Block & on axis & fuel floor & & \\
        Steer & 2 to 9 & direction & 50~$\mu$rad & $1.6 \cdot 10^{-4}$ \\
        Spoof time & in reach & clock, \asm{6} & 1~ms & \\
        Fixed override & below 9 & authentication & & \\
        Adaptive override & below 2.4 & level & $10\%$ spread & $< 10^{-6}$ \\
        Saturate & below 9 & power & $+2$~dB & none \\
        Jam & below 9 & none & & \\
        Use silence & below 1.2 & none, \asm{2} & & \\
        \bottomrule
    \end{tabular*}
\end{table}

\begin{figure}[t]
    \centering
    \includegraphics[width=0.9\columnwidth]{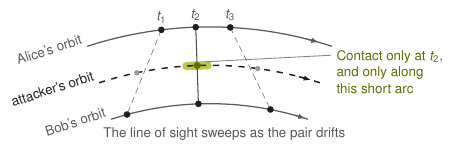}
    \caption{An attacker on an intermediate orbit. Alice and Bob fly at different altitudes, so their line of sight sweeps as the pair drifts. The attacker enters a free circular orbit between them, drawn dashed. The three craft align only at isolated instants, here at $t_2$ but not at $t_1$ or $t_3$. The attacker thus obtains a contact with the two-way seats, in green, of a few minutes per synodic cycle, and the rounds between contacts lock it out.}
    \label{fig:rider}
\end{figure}

\para{Intermediate orbits}
Figure~\ref{fig:rider} shows an attacker on an intermediate orbit. On a crossing pass at 10~km/s relative speed and 1000~km range, the line of sight turns at about 10~mrad/s, and a rigid 1~km standoff then needs about 0.1~m/s$^2$, some 400 times the floor.

\begin{figure}[t]
    \centering
    \includegraphics[width=\columnwidth]{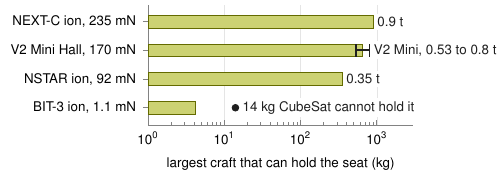}
    \caption{Electric propulsion against the hold requirement of the canonical seat, a mean $2.6 \times 10^{-4}$~m/s$^2$ held continuously through every eclipse. Each green bar is the largest craft a thruster can hold on the seat, its thrust divided by the requirement. The dot marks the 14~kg CubeSat that flies BIT-3 with up to 2.39~km/s, and the range the 0.53 to 0.8~t reported for the Starlink V2~Mini. NSTAR delivers 92~mN at 2.3~kW and NEXT-C 235~mN at 6.9~kW. A CubeSat ion thruster cannot hold its own craft, and gridded ion engines suffice only below about a tonne.}
    \label{fig:propulsion}
\end{figure}

\para{Electric propulsion}
Figure~\ref{fig:propulsion} compares thrusters with the hold requirement~\cite{busek2021bit3, sovey1997nstar, shastry2017nextc, jewett2023v2mini, clark2023v2mini, krebs_v2mini}. The V2~Mini Hall thruster holds the seat at about 80 percent of full thrust on the lightest reported craft. At a specific impulse of 2500~s, a year of the cost on the 1000~km link is 28 percent of the craft's mass. For such craft, the operator should judge the threshold by mass and thrust rather than by the three budget classes.

\section{Ethics Considerations}\label{app:ethics}

This work develops a defensive authentication primitive and quantifies known classes of attack on optical crosslinks. It involves no human subjects, no personal data, and no interaction with live systems. The constellation evaluation uses only public CelesTrak orbital elements, processed offline.

We considered whether the per-seat costs or the finding that short inter-plane links are the cheapest targets could help an attacker. Both follow from public physics, public terminal data, and public orbits, so a capable attacker can already derive them. No deployed product gains a new vulnerability. Operators instead gain a new defensive option, so no disclosure was needed. The main risk is overreliance, since an impostor that breaks \asm{9}, \asm{14}, or \asm{15} can stay accepted (\limref{1}, \limref{2}, \limref{5}).

\section{Artifact Availability}\label{app:artifact}

A Python artifact recomputes the bounds, tables, and propagations, with a harness that checks 182 printed values against the closed forms and the script outputs. It also holds the constellation analysis with its public element snapshot, the inclined-orbit $\symref{J2}{J_2}$ check, the dwell search for intermediate orbits, the channel simulation, the relay of two craft, and the sources of all figures. It is available anonymously at \url{https://anonymous.4open.science/r/FuelasIdentity1/}.

\section{Use of Generative AI}\label{app:ai}
We used Claude by Anthropic through the Claude Code tool. Generative AI was used for editorial purposes in this manuscript, and all outputs were inspected by the authors to ensure accuracy and originality. These edits cover grammar, spelling, wording, and presentation in all sections.

Claude also drafted code, so its use is part of our method. From models, equations, and data that the authors chose, it drafted the artifact scripts behind Sections~\ref{sec:gate} to~\ref{sec:discussion} and the figure sources, which saved implementation time. Drafted code can hide errors. The authors therefore reviewed every script, the harness of Appendix~\ref{app:artifact} checks 182 printed values, and \rqref{3} confirms the floor against numerical propagation and NASA practice. The scripts call no AI model and reproduce every output byte for byte.

\end{document}